\documentclass[final,3p,times,twocolumn]{elsarticle}

\usepackage{epsfig}

\usepackage{amssymb}

\usepackage{caption}
\usepackage{amsfonts}
\usepackage{subfig}
\usepackage{float}
\usepackage{amsmath}
\usepackage{hyperref}
\usepackage{url}

\usepackage{xcolor}

\newcommand{\minew}[1]{{\color{black}{#1}}}
\newcommand{\miold}[1]{\iffalse{#1}\fi}

\usepackage[switch]{lineno}
\newcommand{\newtwo}[1]{{\color{black}{#1}}}
\newcommand{\oldtwo}[1]{\iffalse{#1}\fi}

\biboptions{numbers,sort&compress}

\journal{Biomedical Signal Processing and Control}

\begin{document}

\begin{frontmatter}



\title{A Deep Learning Method for Beat-Level Risk Analysis and Interpretation of Atrial Fibrillation Patients during Sinus Rhythm}

\author[1]{Jun Lei}

\author[1,2]{Yuxi Zhou\corref{cor2}}
\author[1]{Xue Tian}
\author[3]{Qinghao Zhao}
\author[4,5]{Qi Zhang}
\author[6]{Shijia Geng}
\author[1]{Qingbo Wu}
\author[7,8]{Shenda Hong\corref{cor2}}

\cortext[cor2]{Corresponding authors. Email: joy\_yuxi@pku.edu.cn, hongshenda@pku.edu.cn.}

\affiliation[1]{organization={Department of Computer Science},
            addressline={Tianjin University of Technology}, 
            city={Tianjin},
            postcode={300384},
            state={Tianjin},
            country={China}}

\affiliation[2]{organization={DCST, BNRist, RIIT, Institute of Internet Industry},
            addressline={Tsinghua University}, 
            city={Beijing},
            postcode={100084},
            state={Beijing},
            country={China}}

\affiliation[3]{organization={Department of Cardiology},
            addressline={Peking University People’s Hospital}, 
            city={Beijing},
            postcode={100044},
            state={Beijing},
            country={China}}

\affiliation[4]{organization={National Key Laboratory of General Artificial Intelligence},
            addressline={Peking University}, 
            city={Beijing},
            postcode={100871},
            state={Beijing},
            country={China}}

\affiliation[5]{organization={School of Intelligence Science and Technology},
            addressline={Peking University}, 
            city={Beijing},
            postcode={100871},
            state={Beijing},
            country={China}}

\affiliation[6]{organization={HeartVoice Medical Technology},
            city={Hefei},
            postcode={230088},
            state={Anhui},
            country={China}}

\affiliation[7]{organization={National Institute of Health Data Science},
            addressline={Peking University}, 
            city={Beijing},
            postcode={100871},
            state={Beijing},
            country={China}}

\affiliation[8]{organization={Institute of Medical Technology},
            addressline={Peking University}, 
            city={Beijing},
            postcode={100871},
            state={Beijing},
            country={China}}

\begin{abstract}
Atrial Fibrillation (AF) is a common cardiac arrhythmia. Many AF patients experience complications such as stroke and other cardiovascular issues. Early detection of AF is crucial. \newtwo{The great majority of} algorithms can only distinguish ``AF rhythm in AF patients'' from ``sinus rhythm in \miold{normal} \minew{healthy} individuals'' . However, AF patients do not always exhibit AF rhythm, \minew{most of the time they present with sinus rhythms, and there is also a potential risk of AF in sinus rhythms. How to detect AF from sinus rhythm is a challenge.} \miold{posing a challenge for diagnosis when AF rhythm is absent.} To address this, this paper proposes a novel artificial intelligence (AI) algorithm to distinguish ``sinus rhythm in AF patients'' and ``sinus rhythm in \miold{normal} \minew{healthy} individuals'' in beat-level. \minew{We cut 1 second of sinus beats from single-lead Electrocardiogram(ECG) data and fed them to the Net1d model, a deep learning model that processes one-dimensional data, to obtain the risk probability of each beat. Besides, we have also introduced the} beat-level risk interpreters, trend risk interpreters, addressing the interpretability issues of deep learning models and the difficulty in explaining AF risk trends. Additionally, the beat-level information fusion decision is presented to enhance model accuracy. The experimental results demonstrate that the average AUC for single beats used as testing data from CPSC 2021 dataset is 0.7314, \minew{with an average accuracy of 0.6606 and an F1 score of 0.6470.} By employing 150 beats for information fusion decision algorithm, the average AUC can reach 0.7591, \minew{while the average accuracy and F1 score improve to 0.6887 and 0.6749.} Compared to previous segment-level algorithms, we utilized beats as input, reducing data dimensionality and making the model more lightweight, facilitating deployment on portable medical devices. Furthermore, we draw new and interesting findings through average beat analysis and subgroup analysis, considering varying risk levels. Our code is publicly available at \href{https://github.com/leijsen/ECGBeat4AFSinus}{https://github.com/leijsen/ECGBeat4AFSinus} .
\end{abstract}




\end{frontmatter}


\section{Introduction}
\begin{figure*}[t]
    \centering
    \includegraphics[width=1\linewidth]{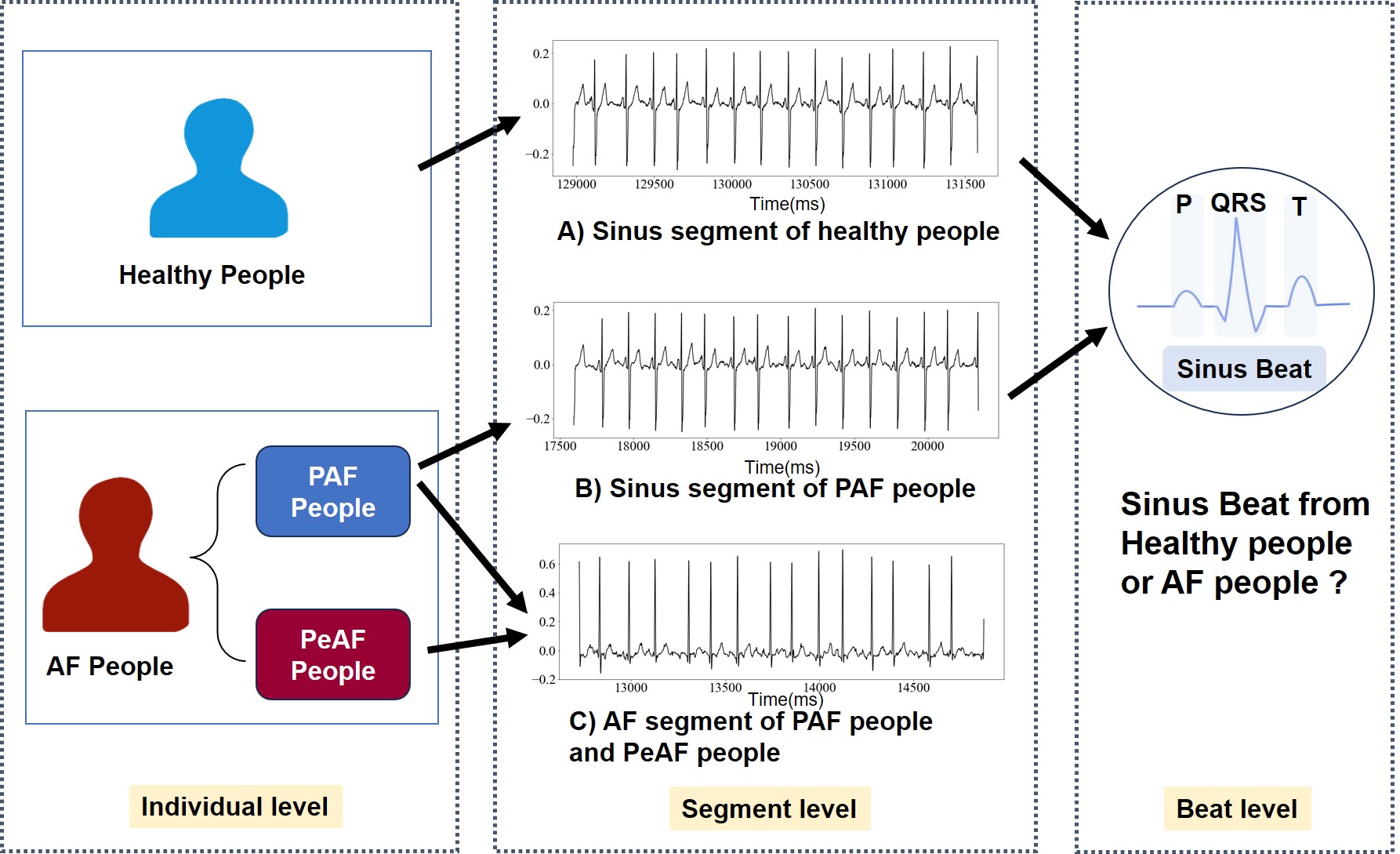}
    \caption{Descriptions of different levels of data. Segment-level data are longer than beat-level data, with any data longer than beat-level considered as a segment. The focus of this paper is to identify sinus beat in healthy people and people with AF.}
    \label{fig:intro}
\end{figure*}

Atrial Fibrillation (AF) is a serious cardiac disease that leads to a significant number of patients developing the condition and facing mortality, yet the diagnostic rate remains low \cite{giannopoulos2023p, attia2019artificial}. Initially presenting as intermittent episodes that spontaneously terminate, AF is a covert disease with an incidence that increases with age \cite{ma2022multistep, bao2022paroxysmal, gunduz2023atrial, wen2022comparative}. As illustrated in Figure \ref{fig:intro}, \minew{There are different levels of AF patients: paroxysmal atrial fibrillation (PAF) and persistent atrial fibrillation (PeAF) \cite{wen2022comparative,an2022percept,wang2022two}. They have different segment types and beat types.} \newtwo{PAF is characterized by intermittent episodes of AF, resulting in discontinuous segment types, where periods of normal sinus rhythm alternate with AF segments. The beat types in PAF show an alternating pattern of normal heartbeats and AF beats. In contrast, PeAF is marked by prolonged and sustained AF episodes, leading to continuous AF segments. Consequently, the beat types in PeAF are predominantly AF beats, with minimal or no presence of normal sinus beats.} Without intervention, PAF may progress to PeAF or even permanent AF, \miold{posing serious harm to human health} \minew{leading to an increased risk of thrombus formation and stroke. Additionally, prolonged untreated AF can result in heart failure and other cardiac-related complications. Therefore, early detection is very important} \cite{ma2022multistep,zhou2019k}.

Electrocardiogram (ECG) is the most commonly used screening tool for AF, but its effectiveness in early diagnosis is limited \cite{perez2019large, guo2019mobile}. Patients in the early stages are mostly PAF \cite{yue2021automatic}. Most of the time, patients with PAF exhibit sinus rhythm. Existing algorithms can only distinguish between ``AF rhythm in AF patients'' and ``sinus rhythm in \miold{normal} \minew{healthy} individuals'', making it challenging to diagnose AF when AF rhythm is absent. \minew{This potential risk of detecting AF in sinus rhythm is what we call the AF substrate.} Recently, \cite{basso2023efficient,saadatnejad2019lstm}the widespread adoption of wearable devices has increased the likelihood of collecting AF rhythm data through long-term ECG monitoring. However, patients are unlikely to wear these devices continuously due to comfort issues and high costs \cite{noseworthy2022artificial}. Additionally, \minew{interpreting AF substrate which is prone to the initiation of AF requires significant amount of expertise \cite{torres2020multi}.} Therefore, establishing an artificial intelligence (AI) algorithm that identifies ``sinus rhythm in AF patients'' and ``sinus rhythm in \miold{normal} \minew{healthy} individuals'' is crucial for preventing further complications and avoiding fatalities.

\minew{In the field of deep learning, current research on AF detection during sinus rhythm is predominantly limited to segment-level algorithms \cite{attia2019artificial}, with no specific algorithms designed for beat-level analysis. The real-life beat waveform is depicted in Figure \ref{fig:intro} and we show the relationship between individual, segment, and beat-level data. In this study, we focused on the differences between P and T waves in healthy people and people with AF. The P wave represents the atrial contraction phase, while the T wave represents the ventricular recovery and repolarization phase \cite{bozyigit2020classification}. \newtwo{However, atrial repolarization also occurs during the T wave, though it is mainly obscured by ventricular repolarization. This interaction might contribute to the T wave’s significance in predictions.} \newtwo{In the meantime,} AF burden also leads to ventricular myocardium remodeling, resulting in T-wave alterations \cite{kortl2022atrial}. Medical research indicates differences in P-waves between sinus rhythm in AF patients and sinus rhythm in healthy individuals \cite{giannopoulos2023p, myrovali2023identifying, martinez2013morphological}, underscoring the feasibility of beat-level algorithms. Compared to segment data, beat-level data offers finer granularity, facilitating more detailed analysis of risk variations. Additionally, beat-level data has lower dimensionality, leading to lighter models that are suitable for deployment on portable medical devices. Moreover, the ability to segment ECG signals allows for more samples from each patient, and decision methods incorporating beat-level information can enhance accuracy.}

Addressing the limitations of existing AI algorithms for beat-level AF detection during sinus rhythm, we have identified the following three challenges. First, analyzing beat-level data is more challenging due to lower information content and increased noise during detection. Second, existing algorithms struggle to analyze dynamic changes in patient risk and lack interpretability. Morphological differences in heart rhythms across different stages are minimal, making it challenging to analyze dynamic changes in AF risk. Third, segment-level algorithms are inadequate for considering risk variations, and beat-level studies have been limited to AF segment analysis without considering \miold{normal} \minew{sinus} segment data.

In response to the aforementioned challenges, we proposes an interpretable framework for AF risk prediction based on the variation of sinus beat probability. The key contributions of our work are summarized as follows:
\minew{
\begin{itemize}
    \item We propose a beat-level AF risk analysis algorithm designed to differentiate between "sinus rhythm in AF patients" and "sinus rhythm in \miold{normal} \minew{healthy} individuals." This approach addresses the challenge of diagnosing AF in patients when AF rhythm data is unavailable. Our algorithm can effectively identify AF in patients, even when the majority of the data presents as sinus rhythm.

    \item We introduce algorithms such as the Beat-level Risk Interpreter (BRI), Beat-level Information Fusion Decision (BID), and Trend Risk Interpreter (TRI). These tools provide clinicians with detailed explanatory information about a patient's ECG status during clinical diagnosis, assisting doctors in making more accurate decisions.
    
    \item We discovered that the average waveforms for patients at different risk levels show that higher risk correlates with a more pronounced disappearance of the T-wave. We present validation results for various patients using the BRI and TRI algorithms, as well as experimental results for the BID algorithm. In subgroup analysis, we found that sinus beats near premature ventricular contractions (BNV) have higher predictive value for AF. Our algorithms also demonstrate improvements in parameter quantity and computational efficiency, making them more suitable for wearable devices.
    
\end{itemize}
}

\section{Related Work}

To accurately identify AF from ECGs, various data input forms have been employed for AF detection. Methods for detecting AF in both AF and \miold{normal} \minew{healthy} data include using segment-level signals as data input for AF detection, using beat-level signals as data input for AF detection, using a combination of single beat-level and segment-level signals as data input for AF detection. There are also methods focused solely on detecting AF in \miold{normal} \minew{sinus} data.

\subsection{Detect AF patients during AF rhythm}

\paragraph{\textbf{Segment-level signals as data input}}
Using segment-level signals as data input for AF detection often involves the application of long short term memory (LSTM) deep learning models and convolutional neural network (CNN)\cite{peimankar2021dens,zhang2022detection,bao2022paroxysmal,jia2022method,huerta2023comparison,bernal2023atrial,jekova2022atrioventricular,ma2022multistep,khurshid2022ecg}. In \cite{peimankar2021dens}, a combination of CNN and  LSTM model is employed to detect waveforms of different heartbeat signals, eliminating the need for feature engineering. While achieving high sensitivity, there is still potential to enhance the capability of filtering AF from other cardiac rhythms. On the other hand, \cite{zhang2022detection} demonstrates the ability to update model parameters and accurately predict AF with different duration and lead distributions, showing better performance in identifying AF from other rhythms. In \cite{ma2022multistep}, a combination of AF rhythm and morphological information improves the accuracy of AF detection, and the algorithm exhibits strong interpretability. However, there is room for improvement in discerning AF from uncertain ECG signals where P-waves may be obscured by noise.

\paragraph{\textbf{Beat-level signal as data input}}
Using beat-level signals as input for AF detection has been explored in \cite{acharya2017deep}. In \cite{acharya2017deep}, a CNN was employed to automatically recognize and classify five different types of heartbeats in ECG signals, including normal (N), supraventricular ectopic (S), ventricular ectopic (V), fusion (F), and unknown (Q) heartbeats. This approach, compared to traditional machine learning methods, has the advantage of automatically learning features from ECG signals without the need for manually designed feature extractors. The paper also addressed the issue of class imbalance in the dataset by using synthetic data to balance the categories of heartbeat data.

\begin{figure*}[!ht]
    \centering
    \includegraphics[width=1\linewidth]{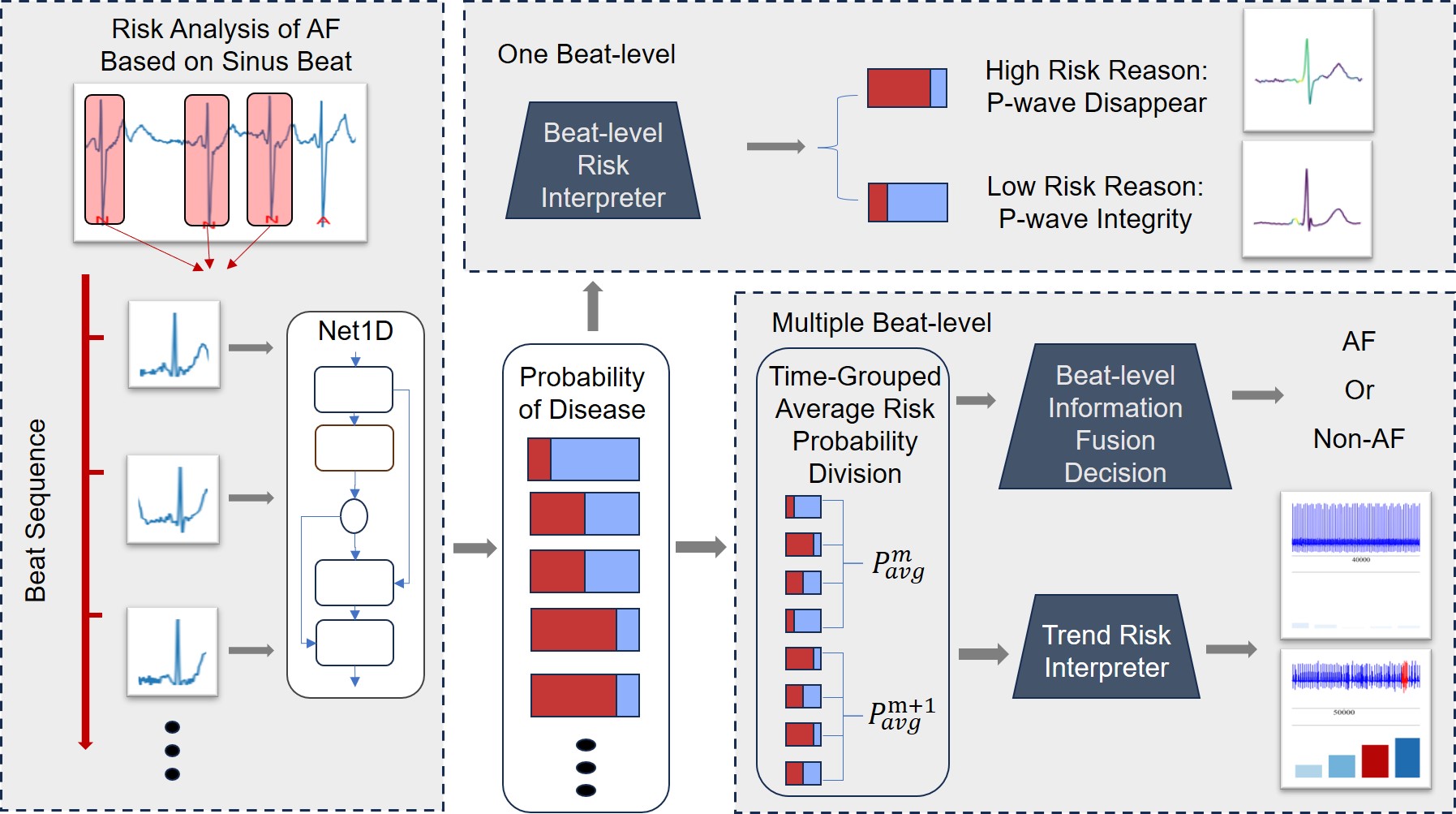}
    \caption{\minew{Overview: First, sinus beats are extracted from the ECG data and input into Net1D model to obtain the probability of being predicted as AF. Based on the number of beats, two processing levels can be selected: for a single beat, the Beat-level Risk Interpreter (BRI) is used for risk interpretation; for a substantial number of beats, the beat-level probabilities are averaged over time groups using the Time-Grouped Average Risk Probability Division (TGD), and the Beat-level Information Fusion Decision (BID) is employed for AF substrate detection. Additionally, the Trend Risk Interpreter (TRI) analyzes trends based on the averaged probabilities over time groups.}}
    \label{overview}
\end{figure*}

\begin{figure*}[!ht]
    \centering
    \includegraphics[width=1\textwidth]{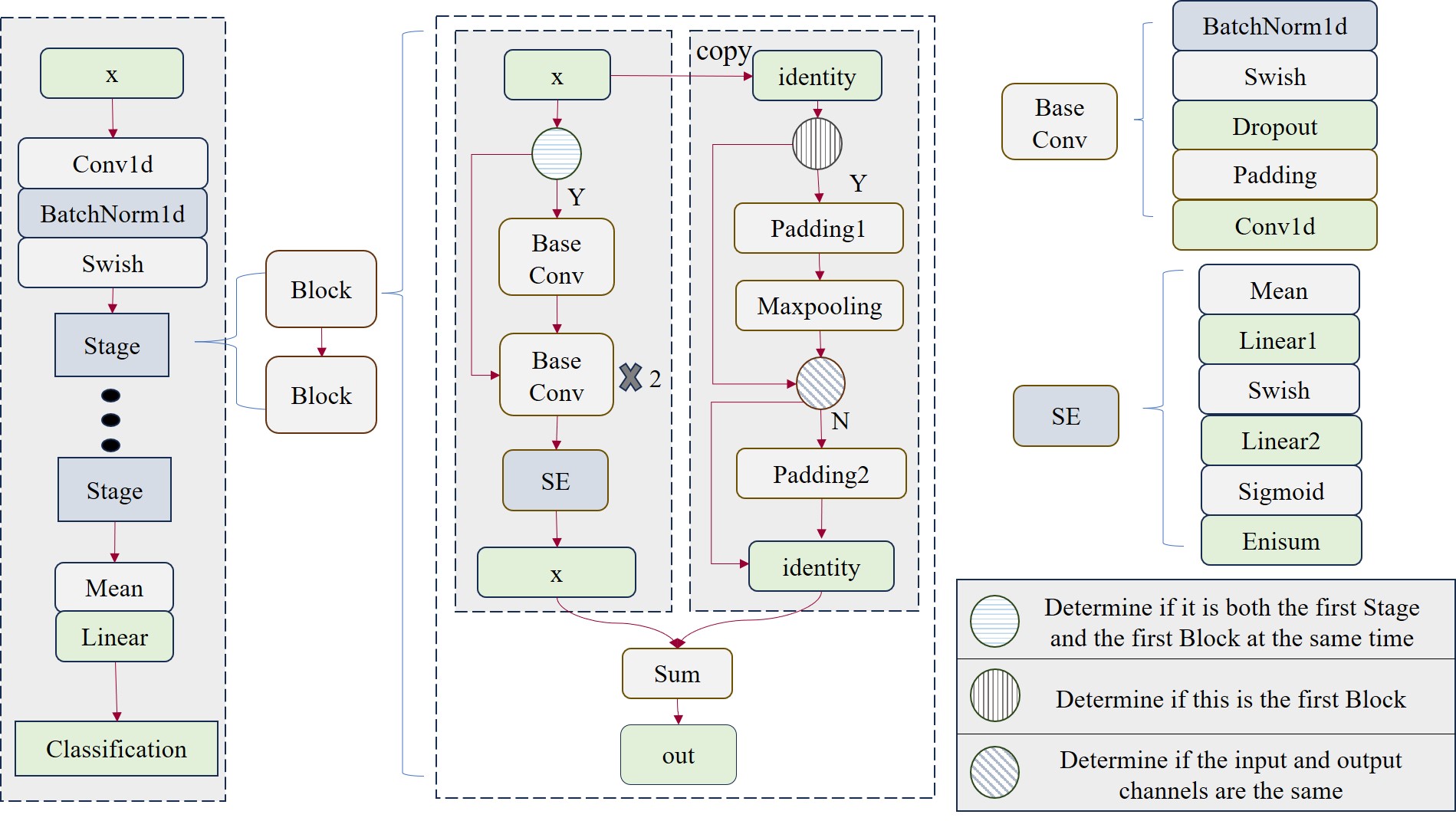}
    \caption{Architecture of Net1d}
    \label{Net1d}
\end{figure*}

\paragraph{\textbf{Beat-level and segment-level signals combined as data inputs}}
Using a combination of beat-level and segment-level signals as input for AF detection has been explored in studies such as \cite{hong2019mina} and \cite{wang2022two}. In \cite{hong2019mina}, MINA combines CNN and bidirectional Long Short-Term Memory network (Bi-LSTM) to extract domain-specific features at different levels (beat-level, rhythm-level, and frequency-level). These features are then combined with ECG data, and a multi-level attention model is employed to enhance the interpretability of the model. This allows the model to identify key beat positions, significant rhythm variations, and important frequency components in ECG signals. \cite{wang2022two} proposes a novel two-step approach for detecting AF events. The first step involves using a Support Vector Machine (SVM) to classify the rhythm type of ECG signals into three categories: non-AF, PAF, and PeAF. The second step utilizes a CNN model to classify heartbeats predicted as AF rhythms, determining the onset and offset points of AF events.

\subsection{Detect AF patients during sinus rhythm}

\newtwo{Only use sinus ECG signal as data input to detect AF has articles \cite{attia2019artificial, dupulthys2024single, kim2024identification, gruwez2023detecting, hygrell2023artificial}}. \cite{attia2019artificial} employed a ResNet architecture to identify ECG features indicative of AF during sinus rhythm. The study designed a method for data collection in interest windows for AF and non-AF patients, allowing the model to distinguish between the sinus rhythm of AF and non-AF patients. \minew{AUC values can reach 0.87-0.90.} However, the collection method may introduce errors, and the discrete nature of temporal sampling could lead to misclassification of certain patients. \minew{ \cite{kim2024identification} compared the effects of RNN, LSTM and Resnet50 \newtwo{using a dataset comprising 13,509 ECGs from 6,719 patients, from which 10,287 normal sinus rhythm ECGs from 5,170 patients were selected.}. Resnet had the best effect, with an accuracy of 65.8\%, an F1 of 71.9\% and an AUC of 0.79. \newtwo{\cite{gruwez2023detecting} detected PAF patients on a sinus dataset consisting of 494,042 ECGs in sinus rhythm from 142,310 patients,} with a high AF prevalence, achieving an accuracy of 78.1\% and an AUC of 0.87. \newtwo{However, they are not comparable with our study because they did not focus on beat-level deep learning methods, which offer finer granularity and potentially enhanced accuracy by analyzing individual heartbeats rather than broader ECG segments.}}

In the four data input forms mentioned for AF detection, none have utilized beat-level sinus rhythm heartbeats as input data. Patients with PAF often exhibit subtle symptoms in the early stages. Therefore, analyzing \miold{abnormal} \minew{unhealthy} signs in sinus rhythm heartbeats in ECG is crucial for preventing AF. Furthermore, comparison with widely used segment-level signal data reveals that beat prediction probabilities fluctuate with the emergence of AF segments. This indicates that beat-level data exhibits higher sensitivity for AF alerts, enabling timely diagnosis and treatment for patients.

\section{Methods}

\minew{In this section, we define the problem and outline the algorithmic process. We detail the Risk Analysis of AF Based on Sinus Beat, where we obtain the probability of disease for each beat. For individual beats, the BRI module interprets risk; for multiple beats, TGD averages the risk probability. BID is used for AF decision-making, and TRI interprets risk trends. Our code is publicly available at \href{https://github.com/leijsen/ECGBeat4AFSinus}{https://github.com/leijsen/ECGBeat4AFSinus}
}

\subsection{Problem definition}

\minew{
A patient has multiple diagnostic segments of ECG, denoted as \( S_i \) for \( i \leq K \), where \( K \) is the number of segments. Each segment consists of two-lead ECG data \( X \in \mathbb{R}^{2 \times N} \), with \( N \) representing the length of each lead signal. We focus on one lead, denoted as \( x \in \mathbb{R}^{1 \times k} \), where \( k \) is the number of R-peaks. Using a sampling rate of 200 Hz and R-peak localization, we extract beat-level data \( d \), representing one heartbeat. According to the doctor's annotation, we extracted the beats marked as 'N' separately.

Given a sinus beat \( d \), our task is to output a binary label \( Y \in \{0, 1\} \), indicating whether the sinus beat is non-AF or an AF segment, along with the risk probability \( p \in [0, 1] \).

We define a continuous time series of probabilities \( P = \{p_1, p_2, \ldots, p_k\} \). The average risk probability for the \( m \)-th time group is \( P_{avg}^m \), where \( m \in \{1, 2, \cdots, \lceil \frac{k}{n} \rceil \} \). \newtwo{Here, \( n = j - i + 1 \) and n is the number of beats in a time group (\( 1 \leq n \leq k \)), where \( i \) represents the \( i \)-th probability value and \( j \) represents the \( j \)-th probability value in the sequence, with \( j \geq i \).} For a sequence of \( t \) average risk probabilities \( \{P_{avg}^1, P_{avg}^2, \cdots, P_{avg}^t\} \), we can interpret variations, providing specific locations of unhealthy ECG features and the level of AF risk. The variable explanations are provided in Table \ref{table:symbols}.
}

\subsection{Overview}

\minew{
As illustrated in Figure \ref{overview}, our approach comprises three main modules: an AF risk analysis algorithm based on sinus beats, a One Beat-level module, and a Multiple Beat-level module. The AF risk analysis algorithm predicts AF for unknown patients' sinus beats. The One Beat-level module, using the Beat-level Risk Interpreter (BRI), provides interpretable analysis for individual sinus beats. The Multiple Beat-level module incorporates Time-grouped Average Risk Probability Division (TGD), Beat-level Information Fusion Decision (BID), and Trend Risk Interpreter (TRI). TGD segments continuous beat probabilities into time groups and calculates the average risk probability for each group; BID fuses information across groups of beats to enhance diagnostic accuracy; TRI interprets dynamic changes in patient risk.
}

\minew{
During training, we construct a sequence of segmented sinus beats and input them into the Net1D model. In the testing phase, given multiple beats \( D = \{d_1, d_2, \cdots, d_i\} \) for a patient, we input these into the trained model to obtain predicted probabilities \( P = \{p_1, p_2, \cdots, p_i\} \). The BRI provides risk interpretation for individual beats. TGD transforms \( P \) into average probabilities \( P_{avg} = \{P_{avg}^1, P_{avg}^2, \cdots, P_{avg}^t\} \). BID offers a more accurate AF diagnosis based on \( P_{avg} \), and TRI provides an analysis of risk changes based on \( P_{avg} \).
}

\begin{table}[!ht]
    \centering
    \begin{tabular}{ll}
    \hline
        \textbf{Symbol} & \textbf{Definition} \\ \hline
        $S$ & ECG diagnostic segment \\ 
        $K$ & Number of segments a patient \\ 
        $X\in \mathbb{R}^{2 \times N}$ & Dual-lead ECG data \\ 
        $N$ & Number of ECG data points \\ 
        $x\in \mathbb{R}^{1 \times N}$ & Data from a single lead \\ 
        $k$ & Number of beats in a lead \\ 
        $d$ & Data of a heartbeat \\ 
        $Y$ & Binary label \\ 
        $p$ & Risk probability \\ 
        $P=\{p_1,p_2,\cdots\}$ & Sequential risk probability \\ 
        $n$ & Number of beats in a group \\
        $P_{avg}^m$ & Average risk probability \\
        $m$ & Present time group \\
        $t$ & Continuous t time groups \\
        $P_{avg}=\{P_{avg}^1,\cdots\}$ & Average sequence \\
        $Rloc$ & R-peak localization \\ 
        $Rclas$ & Beat classification \\ 
        $L$ & Length of a beat \\ 
        $l$ & Left index of a beat \\ 
        $r$ & Right index of a beat \\ 
        $D=\{d_1,d_2,\cdots\}$ & Heartbeat data of a patient \\ 
        $L_{y,p}$ & Model loss function \\
        $\alpha$ & Left index of the time group \\
        $\beta$ & Right index of the time group \\ 
        $layer$ & Output layer \\
        $f_l$ & Function of the $l$-th layer \\
        $\theta_l$ & Parameters of the $l$-th layer \\
        $v$ & Feature map value \\
        $f$ & Foward propagation function \\
        $w$ & Output layer weights \\
        $cam$ & Class activation map \\
        $map$ & Mapped result \\
        $pre$ & Predicted label \\
        $threshold$ & Risk threshold \\
        \hline
    \end{tabular}
    \caption{Table of symbols}
    \label{table:symbols}
\end{table}

\subsection{Risk Analysis of AF Based on Sinus Beat}

\minew{
Firstly, we preprocess the selected dataset by applying necessary filtering steps to eliminate baseline drift and optimize signal quality, thereby preventing shortcut issues \cite{geirhos2020shortcut}. Based on doctors' localization of the R-peaks, denoted as $Rloc$, where $Rloc[i]$ represents the time scale of the R-peak for the \(i\)-th heartbeat in the entire ECG (\(1 \leq i \leq k\)), and the labels for each beat, denoted as $Rclas$, where $Rclas[i]$ indicates the type of the \(i\)-th beat (\(1 \leq i \leq k\)), we proceed with segmentation. Using a sampling rate of 200Hz, the length \(L\) of a beat is defined as 200. For the \(i\)-th beat, we define its left index \(l_i\) and right index \(r_i\) as follows:
\begin{equation}
    l_i = Rloc[i] - \frac{L}{2}, \quad r_i = Rloc[i] + \frac{L}{2}
\end{equation}
}

\minew{
All data in a single lead is segmented into beat-level sequences of equal length, denoted as \(D = \{d_1, d_2, \cdots, d_k\}\). Segmentation begins from the 10th R-peak and continues until the 5th R-peak from the end to avoid initial and terminal noise. Beats are classified based on $Rclas$ into either 'N' or non-'N'. 'N'-type beats are retained, while non-'N'-type beats are discarded, resulting in sinus rhythm data for a patient. For binary classification, if a patient has at least one segment \(S_i\) marked as unhealthy by a doctor, all segments \(\{S_1, S_2, \cdots, S_K\}\) for that patient, as well as the beat data sequence \(D\) within each segment, are labeled as 1. Unhealthy segments include PAF or PeAF. Otherwise, all heartbeat data \(D\) is labeled as 0.
}

\minew{
After segmenting the ECG signals into beats, we employ the Net1d model \cite{hong2020holmes} with sinus beat data \(d\) as input to learn implicit AF risk information and determine the presence of AF in the subject. Figure \ref{Net1d} illustrates the architecture of the Net1d model. Net1d conducts feature extraction on the data, involving conv layers, BatchNorm (BN), and swish activation layers. The model comprises 7 stage modules of different shapes, followed by averaging and linear layers to derive probabilities for predicting 0 and 1. The construction of Net1d is straightforward, with a modular structure enabling independent adjustment of hyperparameters in each module. Table \ref{table:modelPara} describes the meaning and values of the parameters of the model. The hyperparameters used for training are shown in Table \ref{table:hyperpara}. For detailed implementation guidelines, please consult the code available at \href{https://github.com/leijsen/ECGBeat4AFSinus}{https://github.com/leijsen/ECGBeat4AFSinus}.}

\begin{table*}[!ht]
    \centering
    \resizebox{1\textwidth}{!}{
        \begin{tabular}{ccc}
            \hline
            Model parameter & Description & Value \\ \hline
            Input channel & Number of input channels & 1 \\ 
            Base filters & Number of output channels in the first 1D convolution module & 32 \\ 
            Ratio & Multiplier for the number of output channels in each module & 1.0 \\ 
            Filter list & Number of output channels for each module & [16, 32, 32, 40, 40, 64, 64] \\ 
            Block list & Number of blocks in each stage module & [2, 2, 2, 2, 2, 2, 2] \\ 
            Kernel size & Size of the convolutional kernel & 8 \\ 
            Stride & Stride of the convolutional kernel & 1 \\ 
            Groups width & Width of the groups for dividing the input channels & 4 \\  
            Classes & Number of output classes & 2 \\ \hline
        \end{tabular}
    }
    \caption{Selection of model parameters.}
    \label{table:modelPara}
\end{table*}

\begin{table}[!ht]
    \centering
    \begin{tabular}{ll}
    \hline
        Hyperparameter & Value \\ \hline
        Learning rate & 1.0E-05 \\
        Epoch & 100 \\
        Batch-size & 32 \\
        Beat-data & 200 \\ \hline
    \end{tabular}
    \caption{Selection of Hyperparameters. Beat data is used to represent the number of data points in a beat-level data.}
    \label{table:hyperpara}
\end{table}

\miold{
Each stage module contains 2 block layers, and a block layer is generally composed of 3 decision points, a baseConv layer, and a SqueezeAndExcitation (SE) layer. The 3 decision points are: checking if this block belongs to both the first stage and the first block, if so, the input goes through the first baseconv layer and then the following two baseconv layers; checking if this block belongs to the first block, in which case, intermediate data goes through padding and maxpool layers; checking if the dimensions of input and output channels are the same, if not, the data goes through another padding layer. Note: the two padding layers have some differences in their processing. The final output is obtained by using the Sum function.
}

\miold{
The specific baseConv layer is composed of BN layer, swish activation layer, dropout layer, padding layer, and conv layer. The detailed SE is composed of averaging layer, linear layer, swish activation layer, linear layer, sigmoid layer, and the enisum layer, defined as follows: considering two matrices, one with dimensions $(a, b, c)$, and the other with dimensions $(a, b)$. Their multiplication, represented by the Einstein summation convention, is given by the formula (\ref{Enisum}).
\begin{equation}
    A_{a*b*c} = B_{a*b*c} \cdot C_{a*b}
    \label{Enisum}
\end{equation}
Where $A, B, C$ represent matrices, and $a$, $b$, and $c$ are the dimensions of the first matrix, and $a, b$ are the dimensions of the second matrix.
}  

\minew{
The cross-entropy loss function \(L_{y, p}\) for the model is defined as:
\[
L_{y, p} = -\left(y \cdot \log(p) + (1 - y) \cdot \log(1 - p)\right)
\]
Here, \(y\) represents the true label, and \(p\) represents the predicted label.
}

\subsection{One Beat-level Module}

\paragraph{Beat-level Risk Interpreter}
In clinical judgment, physicians often focus on the pathological regions of beats \cite{wang2023ecggan}. Therefore, we utilize the Class Activation Map (CAM) method to visualize and interpret the model's output. The CAM method highlights which parts of the beat data contribute more to the prediction output. Higher scores result in more vibrant colors, indicating a higher importance of each region for the predicted category. As illustrated in Figure \ref{overview} on the upper right, brighter areas suggest that the model pays more attention to those regions, providing guidance on why the model predicts a particular beat as AF or non-AF. The detailed process is as follows:

\begin{equation}
    layer = f_l(d; \theta_l)
\end{equation}
$layer$ represents the model's output layer, $f_l$ represents the forward propagation function of the $l$-th layer of the model, $d$ represents the input beat-level data, and $\theta_l$ represents the parameters of the $l$-th layer of the model.
\begin{equation}
    v=f(d; layer)
\end{equation}
$v$ represents the feature map value, and $f$ denotes the forward propagation function of the model.
\begin{equation}
    w = \frac{\partial L_{y,p}}{\partial layer}
\end{equation}
$w$ represent the weights of the output layer, and $L_{y,p}$ denotes the loss function.
\begin{equation}
    cam=v \cdot w
\end{equation}
$cam$ represents the class activation map.
\begin{equation}
    map=ReLU(\sum_{i=1}^n cam_i)
\end{equation}
$map$ represents the result after mapping the class activation map, $\sum_{i=1}^n cam_i$ represents the summation across the first dimension of $cam$.

\subsection{Multiple Beat-level Module}

\paragraph{\textbf{Time-Grouped Average Risk Probability Division}}
Given a patient's beat sequence $D$, we obtain the model Net1d's predicted risk probability sequence $P = \{p_1, p_2, \cdots, p_k\}$, where $k$ is the number of beats in the sequence $D$. We divide $n$ consecutive beats into a time group, and the average risk probability $P_{avg}^m$ for the $m$-th time group is calculated as follows:
\begin{equation}
    P_{avg}^m = \frac{\sum_{i=\alpha}^{\beta} p_i}{n}
\end{equation}
Here, $\alpha$ represents the left index of the $n$ consecutive beats, $\beta$ represents the right index, and $m$ is the index of the current time group.
After calculating the average risk probability for each time group, we form the sequence of time group average risk probabilities $P_{avg} = \{P_{avg}^1, P_{avg}^2, \cdots, P_{avg}^t\}$, $t=\lceil\frac{k}{n}\rceil $.

\begin{figure*}[!bt]
\centering
\subfloat[]{\includegraphics[width=0.25\textwidth]{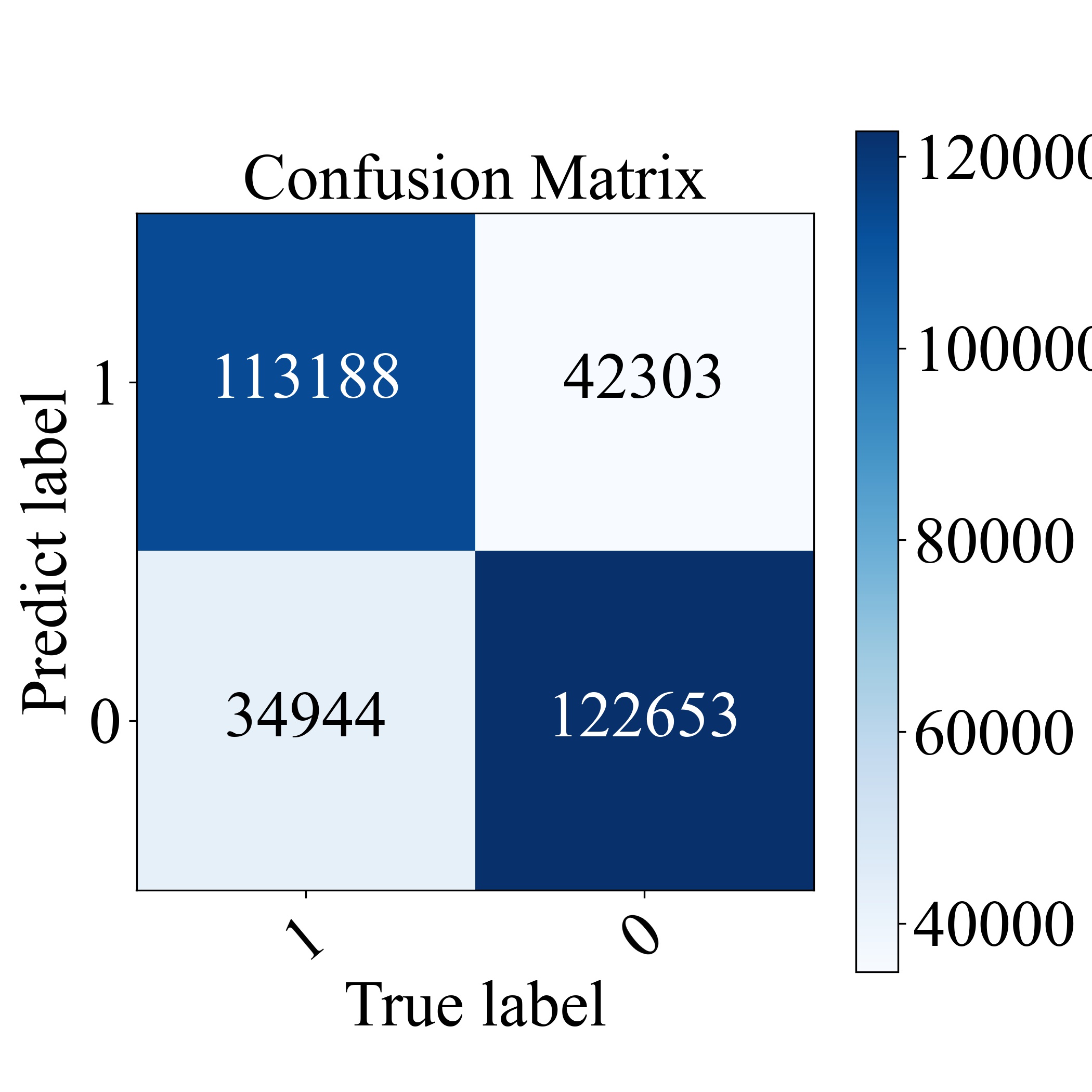}}
\hfill
\subfloat[]{\includegraphics[width=0.25\textwidth]{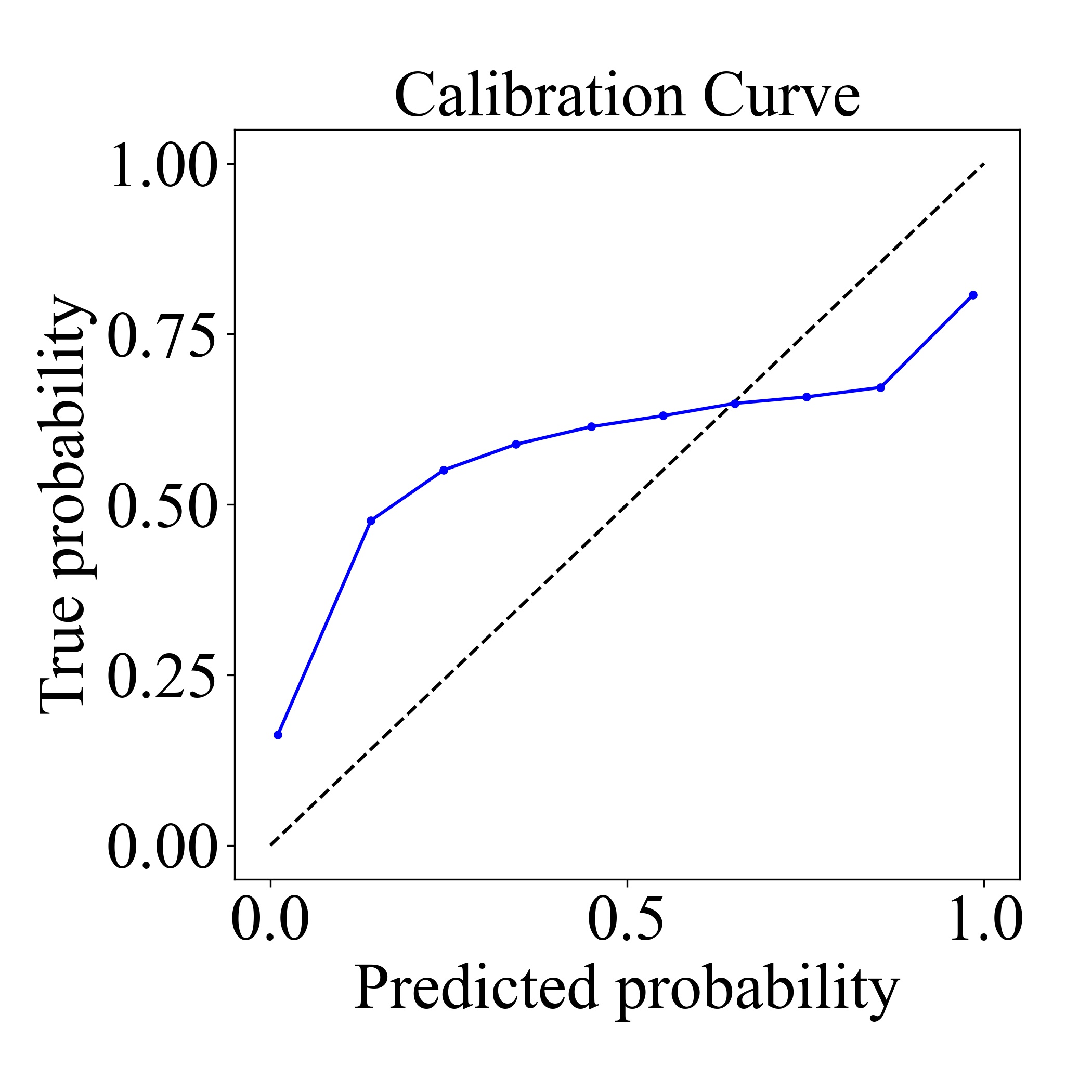}}
\hfill
\subfloat[]{\includegraphics[width=0.25\textwidth]{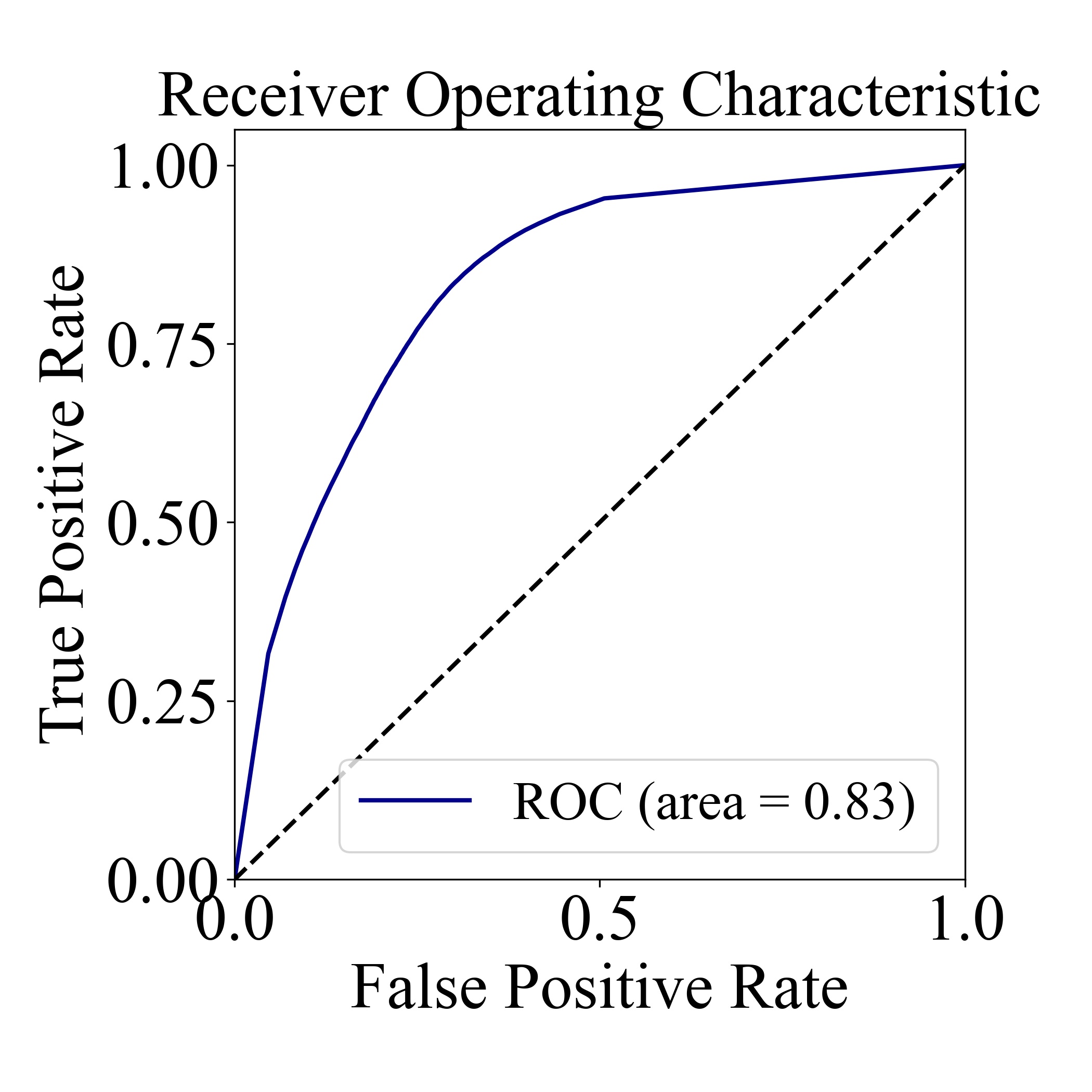}}
\hfill
\subfloat[]{\includegraphics[width=0.25\textwidth]{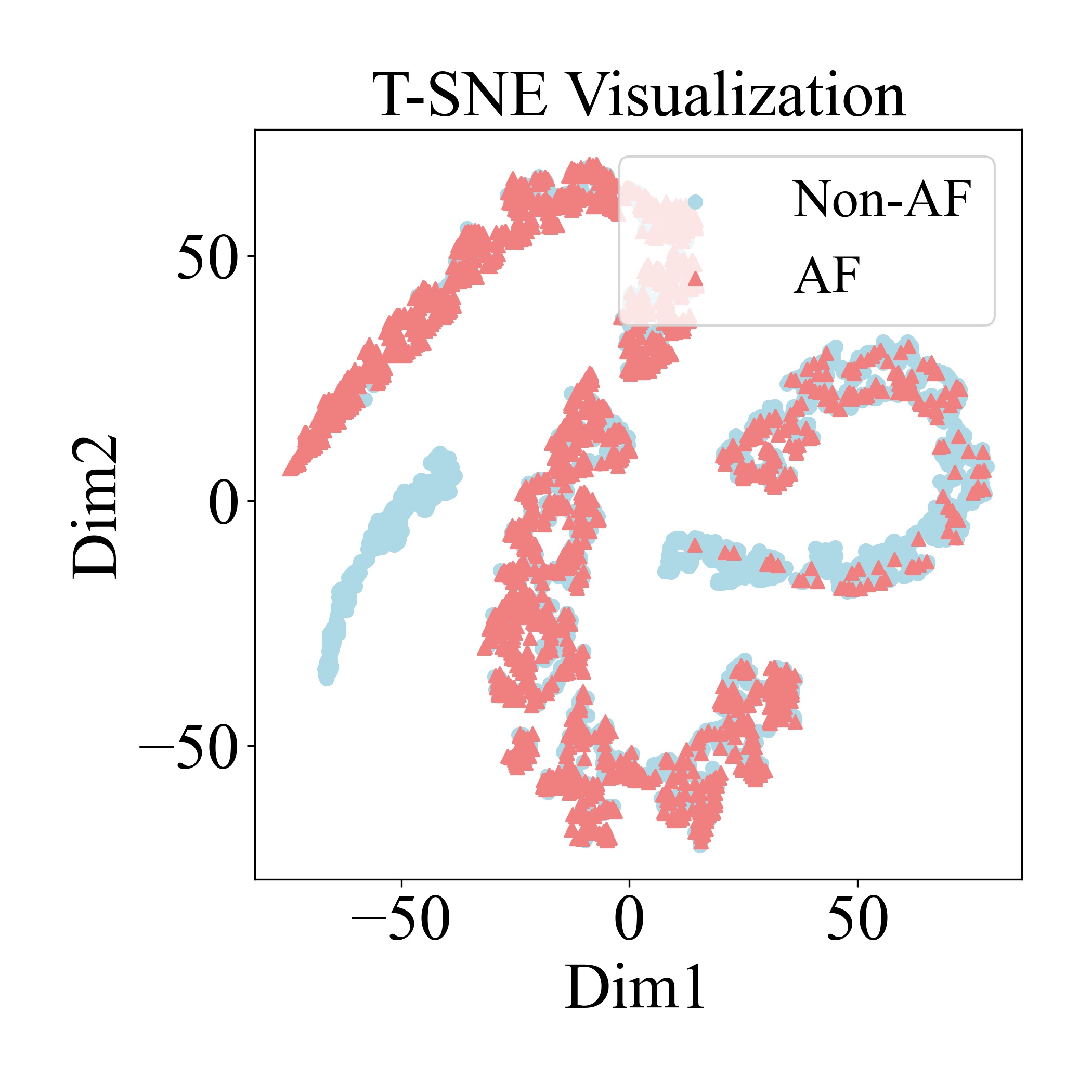}}
\caption{a) Confusion matrix. b) Calibration Curve for AF. c) ROC-AUC Curve. d) T-SNE Clustering of Images.}
\label{fig:performance}
\end{figure*}

\paragraph{\textbf{Beat-level Information Fusion Decision}}
In clinical practice, making a diagnosis based on a single beat-level data may have some randomness. Therefore, leveraging the flexibility and granularity advantages of beat-level data, we propose the BID algorithm. It is described as follows:

We take the average of the predicted probabilities of $n$ consecutive beats for a patient, denoted as $P_{avg}^m$, as the average risk probability for this segment of data. When the average risk probability exceeds a certain threshold, it indicates that the patient has a high probability of having AF.
\begin{equation}
    \text{pre} = 
    \begin{cases}
    1 & \text{if } p_{avg}^m \geq threshold \\
    0 & \text{if } p_{avg}^m < threshold
    \end{cases}
\end{equation}

Where $pre$ represents the predicted label for the patient, with 1 indicating AF and 0 indicating non-AF. $threshold$ denotes the risk threshold defined by the doctor, with $0 \leq threshold \leq 1$, and the risk threshold can be adjusted according to the actual situation.

\paragraph{\textbf{Trend Risk Interpreter}}
In clinical diagnosis, medical practitioners focus extensively on segments displaying \miold{abnormal} \minew{unhealthy} beats, facilitating a thorough analysis of these anomalous cardiac rhythms. Following the acquisition of a sequence of average risk probabilities $P_{avg}$, we undertake an adjusted examination of this $P_{avg}$ sequence to observe the trend of risk variations, as depicted in the lower right corner of Figure \ref{overview}. During testing, when $P_{avg}^m$ falls outside an AF segment, it is depicted in blue. The intensity of blue signifies a higher risk value of $P_{avg}^m$, with deeper shades indicating increased risk. Conversely, when $P_{avg}^m$ corresponds to an AF segment, it is depicted in red. In \miold{normal} \minew{sinus} ECG, each $P_{avg}^m$ value typically remains below the risk threshold $threshold$ and exhibits a stable trend, signifying a low likelihood of abnormality in this beat segment. In contrast, for ECGs of AF patients, $P_{avg}^m$ demonstrates an ascending trend from \miold{normal} \minew{sinus} to AF segments, indicating a heightened probability of abnormality in this segment.

\section{Experiments and results}

In this section, we will first introduce the dataset used, then describe the performance of the model. We will showcase the results of subgroup analysis, comparing the parameters count of models at different levels. Finally, we will highlight interpretability and new discoveries. We primarily utilize the model with index=4 to showcase the results of various experiments.

\begin{table}[!ht]
    \centering
    \resizebox{1\linewidth}{!}{
        \begin{tabular}{cccccc}
            \hline
            \textbf{Index} & \textbf{Accuracy} & \textbf{Recall} & \textbf{Precision} & \textbf{F1} & \textbf{AUC} \\ \hline
            \textbf{1} & 0.5581 & 0.5778 & 0.4836 & 0.5265 & 0.5845 \\ 
            \textbf{2} & 0.6093 & 0.6713 & 0.5426 & 0.6001 & 0.6737 \\ 
            \textbf{3} & 0.7438 & 0.7240 & 0.6693 & 0.6956 & 0.8129 \\ 
            \textbf{4} & 0.7520 & 0.7263 & 0.7630 & 0.7442 & 0.8334 \\ 
            \textbf{5} & 0.6399 & 0.5662 & 0.8155 & 0.6684 & 0.7527 \\ 
            \textbf{Avg} & 0.6606 & 0.6531 & 0.6548 & 0.6470 & 0.7314 \\ \hline
        \end{tabular}
    }
    \caption{Cross-validation results. The indices below represent the test set names for the 5 cross-validation experiments and the averaged results after experiments.}
    \label{table:cross}
\end{table}

\subsection{Dataset}

This study utilized publicly available datasets provided by CPSC 2021, accessible at \cite{wang2021paroxysmal}. These databases consist of 1436 ECG recordings from 105 subjects, selected from the I-lead and II-lead of long-term dynamic ECG signals. In this paper, only the data from the I-lead was used. The duration of the records varies, ranging from 0.14 to 411.11 minutes, with an average duration of 20.33 minutes \cite{wang2022two}. \newtwo{The signals in the dataset were filtered using a bandpass filter with a frequency range of 0.5 Hz to 50 Hz and a sampling frequency of 200 Hz. Specifically, we employed an Infinite Impulse Response (IIR) filter method from the MNE library. By default, this method uses a 4th-order Butterworth filter, which balances the need for effective signal processing with computational efficiency. This filter removes noise and preserves the important features of the ECG signals within the specified frequency range.}

In terms of data processing, this study employed a specific method to label patients with AF. Specifically, if there is at least one instance of AF in the data segment for a given patient, the entire patient is labeled as having AF. Following this classification criterion, a total of 54 subjects were marked as having AF, while the remaining 51 were classified as non-AF patients. This classification was implemented to ensure the accuracy of the model in detecting patients with AF substrate and to enhance the model's sensitivity to AF data.

\begin{figure}[!ht]
    \centering
    \includegraphics[width=1.0\linewidth]{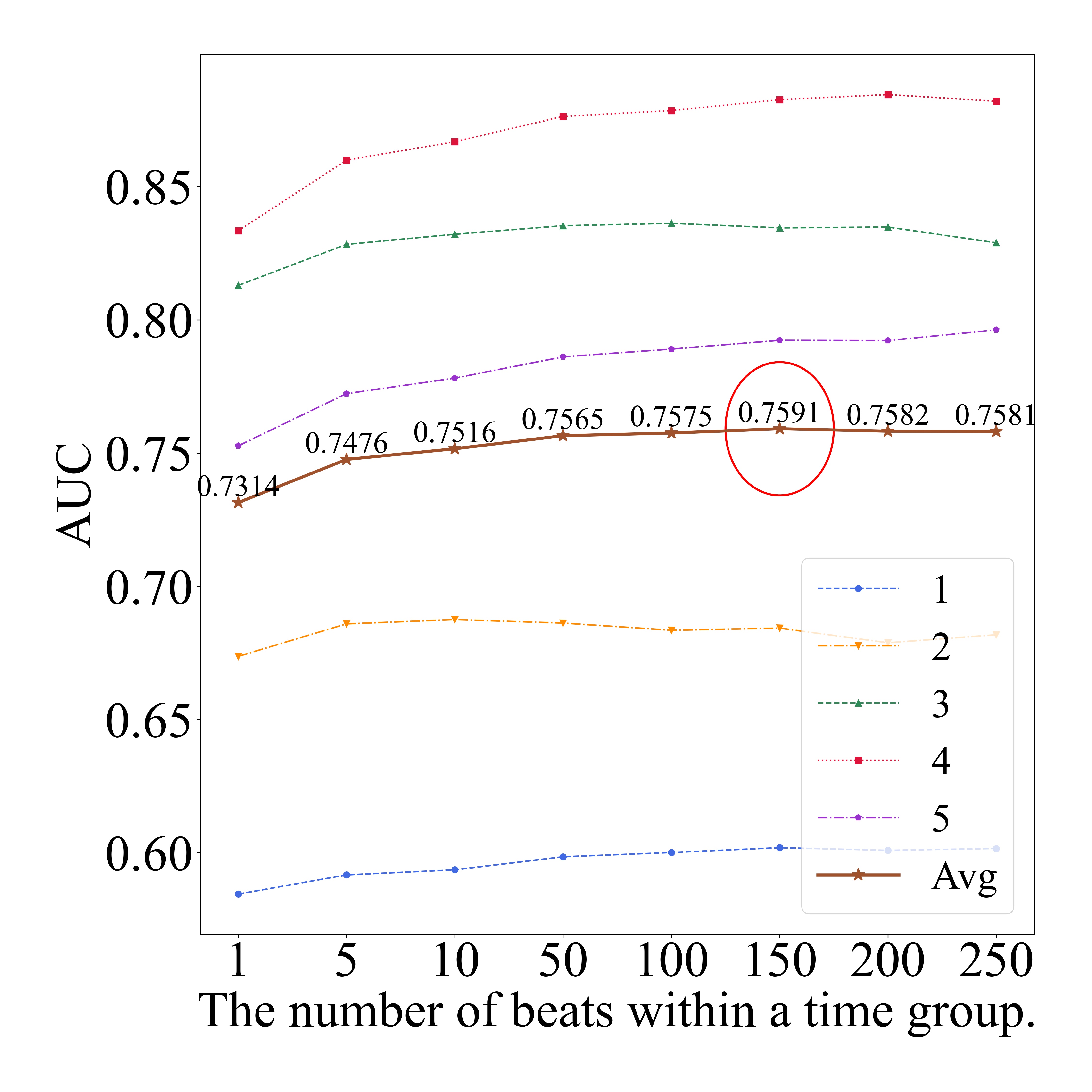}
    \caption{Performance of the beat-level information fusion decision (BID) algorithm. The horizontal axis represents the number of beats in a time group, denoted as $n$. The vertical axis represents the AUC value. Labels indicate different indices corresponding to the index.}
    \label{fig:compareAuc}
\end{figure}

\begin{table}[!ht]
    \centering
    \resizebox{1\linewidth}{!}{
        \begin{tabular}{cccccc}
        \hline
        \textbf{Index} & \textbf{Accuracy} & \textbf{Recall} & \textbf{Precision} & \textbf{F1} & \textbf{AUC} \\ \hline
        \textbf{1} & 0.5608 & 0.5889 & 0.484 & 0.5314 & 0.6019 \\ 
        \textbf{2} & 0.6185 & 0.6739 & 0.5517 & 0.6067 & 0.6843 \\ 
        \textbf{3} & 0.7857 & 0.7325 & 0.7230 & 0.7277 & 0.8345 \\ 
        \textbf{4} & 0.8203 & 0.8235 & 0.8284 & 0.8260 & 0.8826 \\ 
        \textbf{5} & 0.6584 & 0.5796 & 0.8302 & 0.6826 & 0.7923 \\ 
        \textbf{Avg} & 0.6887 & 0.6797 & 0.6835 & 0.6749 & 0.7591 \\ \hline
        \end{tabular}
    }
    \label{table:150BID}
    \caption{The results of BID algorithm uses 150 beats.}
\end{table}

Additionally, we segmented approximately 21,469,915 heartbeats from these records, including 888,067 AF-related data and 1,258,848 \miold{normal} \minew{sinus} heartbeat data.

\begin{figure*}[!ht]
\centering
\subfloat[]{\includegraphics[width=0.2\textwidth]{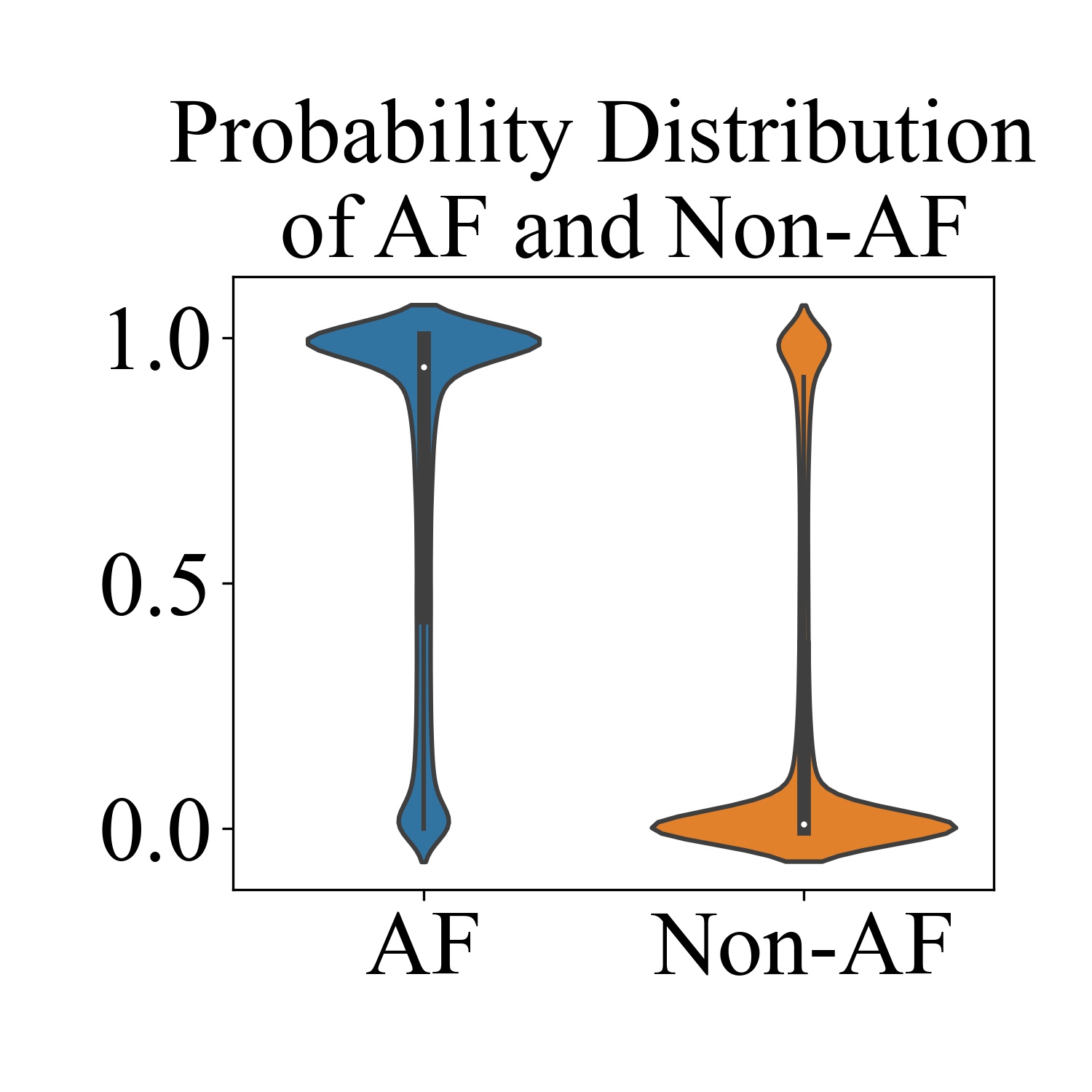}}
\hfill
\subfloat[]{\includegraphics[width=0.2\textwidth]{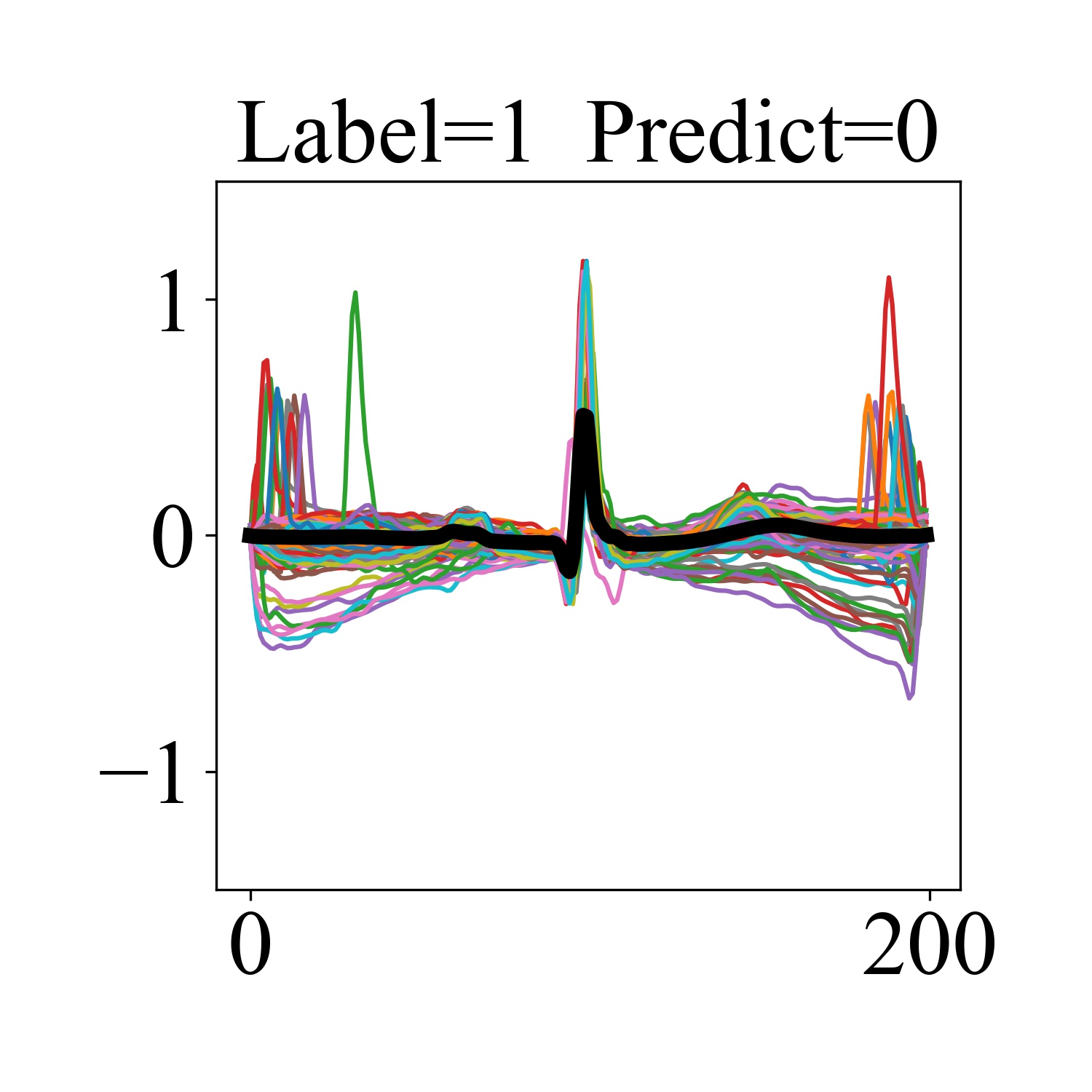}}
\hfill
\subfloat[]{\includegraphics[width=0.2\textwidth]{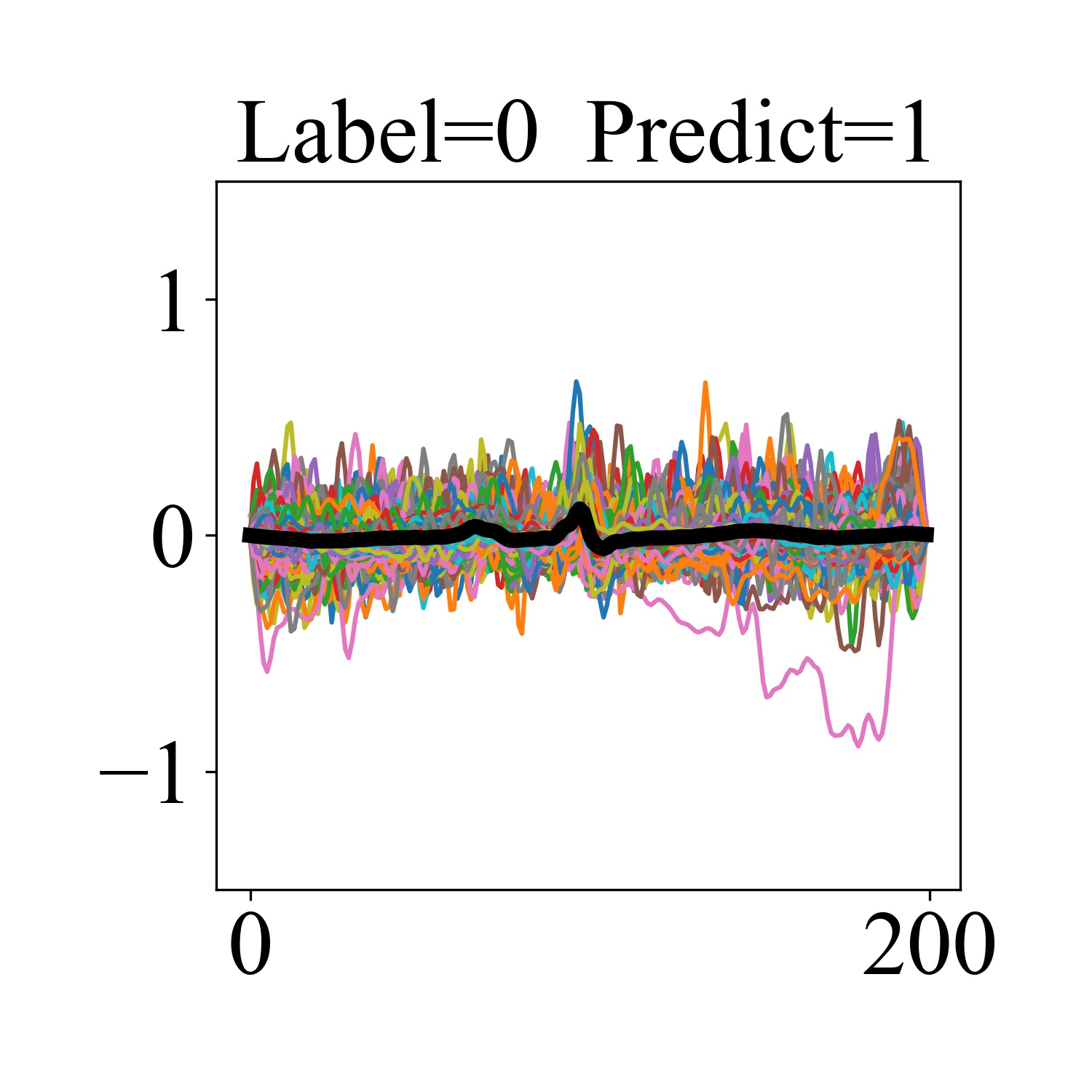}}
\hfill
\subfloat[]{\includegraphics[width=0.2\textwidth]{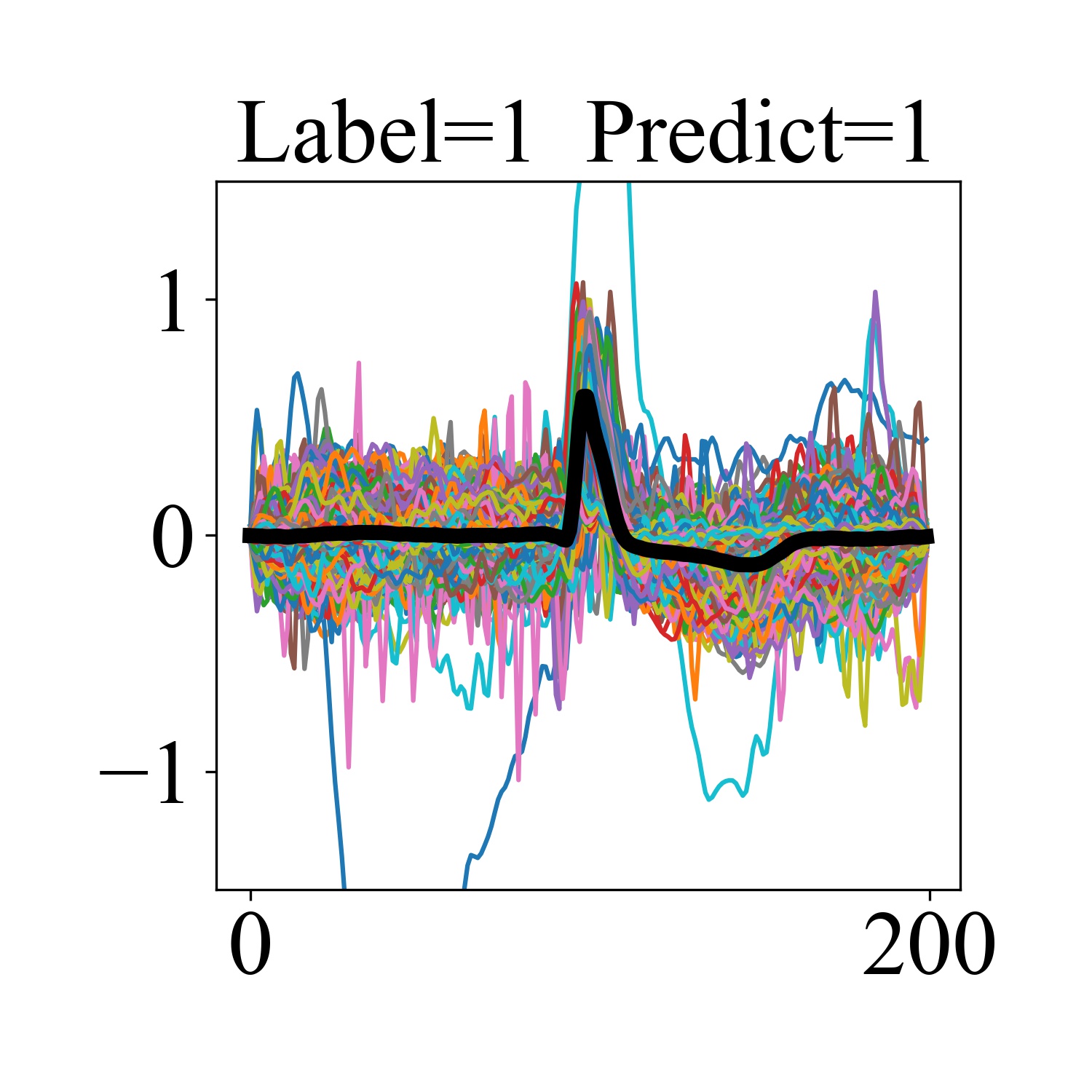}}
\hfill
\subfloat[]{\includegraphics[width=0.2\textwidth]{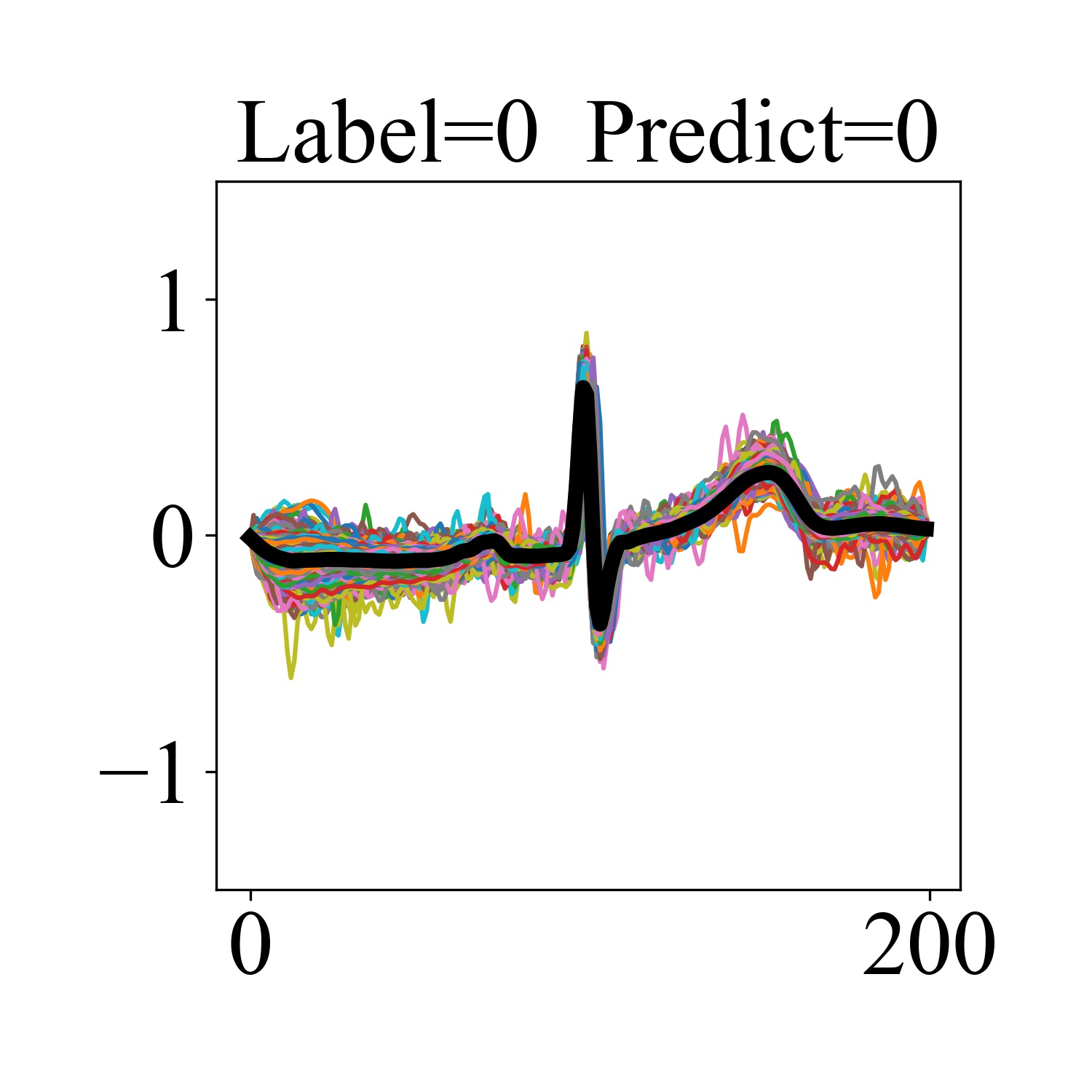}}
\caption{Variations in model prediction probabilities. (a) The horizontal axis represents the groundtruth of the samples (AF or non-AF), while the vertical axis represents the probability assigned by the model for each sample to be AF. The width of the graph represents the relative density of the sample count corresponding to the current probability. (b) \& (c) Illustrate the shapes of samples where the model predicted incorrectly for AF and non-AF labels. The horizontal axis represents the data points of a beat (fixed at 200), and the vertical axis represents the filtered beat values. The bold black waveform represents the result of averaging the values (d) \& (e) Illustrate the shapes of samples where the model predicted correctly for AF and non-AF labels.}
\label{violin}
\end{figure*}

\begin{figure}[t]
    \centering
    \includegraphics[width=1.0\linewidth]{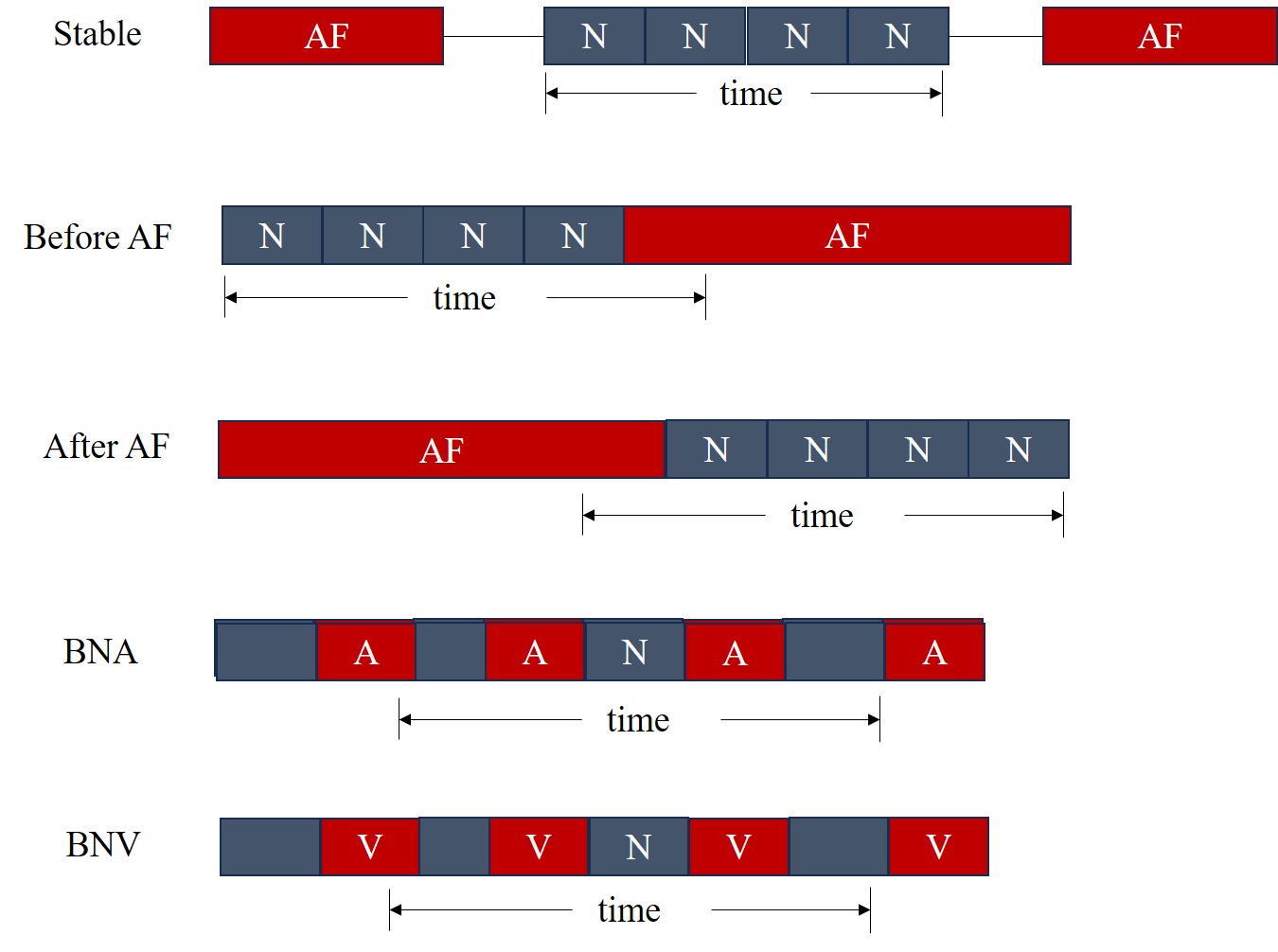}
    \caption{The various partitioning methods for sub-group analysis: Stable, Before AF, After AF, Sinus beat near atrial premature contraction (BNA), Sinus beat near Premature Ventricular Contraction (BNV). The red AF denotes AF segments, the red A represents atrial premature contraction, \newtwo{the red V represents premature ventricular contraction} and the black N represents sinus beat.}
    \label{fig:subgroup}
\end{figure}

\subsection{Classification performance}

We evaluate the model performance using key performance metrics including accuracy, precision, recall, and Area Under Curve (AUC). The results of the 5-fold cross-validation are shown in Table \ref{table:cross}. \minew{Index represents the index of 5 cross-validation, and we show the effect of all to reduce the randomness of the results.} We choose the model with index=4 for displaying the AUC plot in Figure \ref{fig:performance}. The BID enhances the precision of patient prediction, as depicted in Figure \ref{fig:compareAuc}, illustrating the variation of AUC values with different lengths of time group within beat quantity $n$. The graph indicates that as $n$ increases, the AUC value tends to grow, especially in the range from 1 to 10, showing a noticeable improvement. It is evident that, with the assistance of the BID algorithm, beat-level risk assessment can enhance accuracy. When there is a relatively larger number of beats, a higher level of detection can be achieved. However, when the number of heartbeats increases to a certain threshold, further increases can lead to a decrease in results.

We designed Figure \ref{violin} to showcase the probability distribution of model outputs. It illustrates the confidence of the model on different samples and reveals the uncertainty of the model across different categories. By presenting multiple data points and their corresponding average waveforms, we demonstrate both samples predicted incorrectly and those predicted correctly by the model, along with their approximate shapes. The experiment confirms that the majority of beat data is correctly classified. However, there is a small portion that is misclassified, such as beats from AF patients being classified as \miold{normal} \minew{healthy}. This misclassification may occur when certain key features of the beat, such as the P-wave or T-wave, have minor fluctuations, misleading the model into thinking it belongs to a \miold{normal} \minew{healthy} individual. Similarly, some beats from \miold{normal} \minew{healthy} individuals may be misclassified as AF. This could be attributed to the prominent P-wave features of the beat but with a partial disappearance of the T-wave, leading to misclassification.

We conducted a detailed analysis of the model's classification results through the confusion matrix shown in Figure \ref{fig:performance}. This matrix visually displays the model's performance on true positives, false positives, true negatives, and false negatives. The darker the values on the main diagonal, the higher the probability that the model predicted correctly.

In order to further evaluate the calibration performance of the model, we plotted the calibration curve shown in Figure \ref{fig:performance}. This curve illustrates the accuracy of the model's probability predictions. From the curve, it can be observed that the model predictions are overly confident, with predicted probabilities tending to be either 0 or 1, resulting in an inverted sigmoid shape for the blue curve.

Through the T-SNE clustering analysis in Figure \ref{fig:performance}, we observed the distinctiveness of the model. After the model distinguishes the data, there is a clear separation, proving that the model has a certain ability to differentiate between data with different labels.

\begin{figure*}[!tp]
\centering
\subfloat[]{\includegraphics[width=0.2\textwidth]{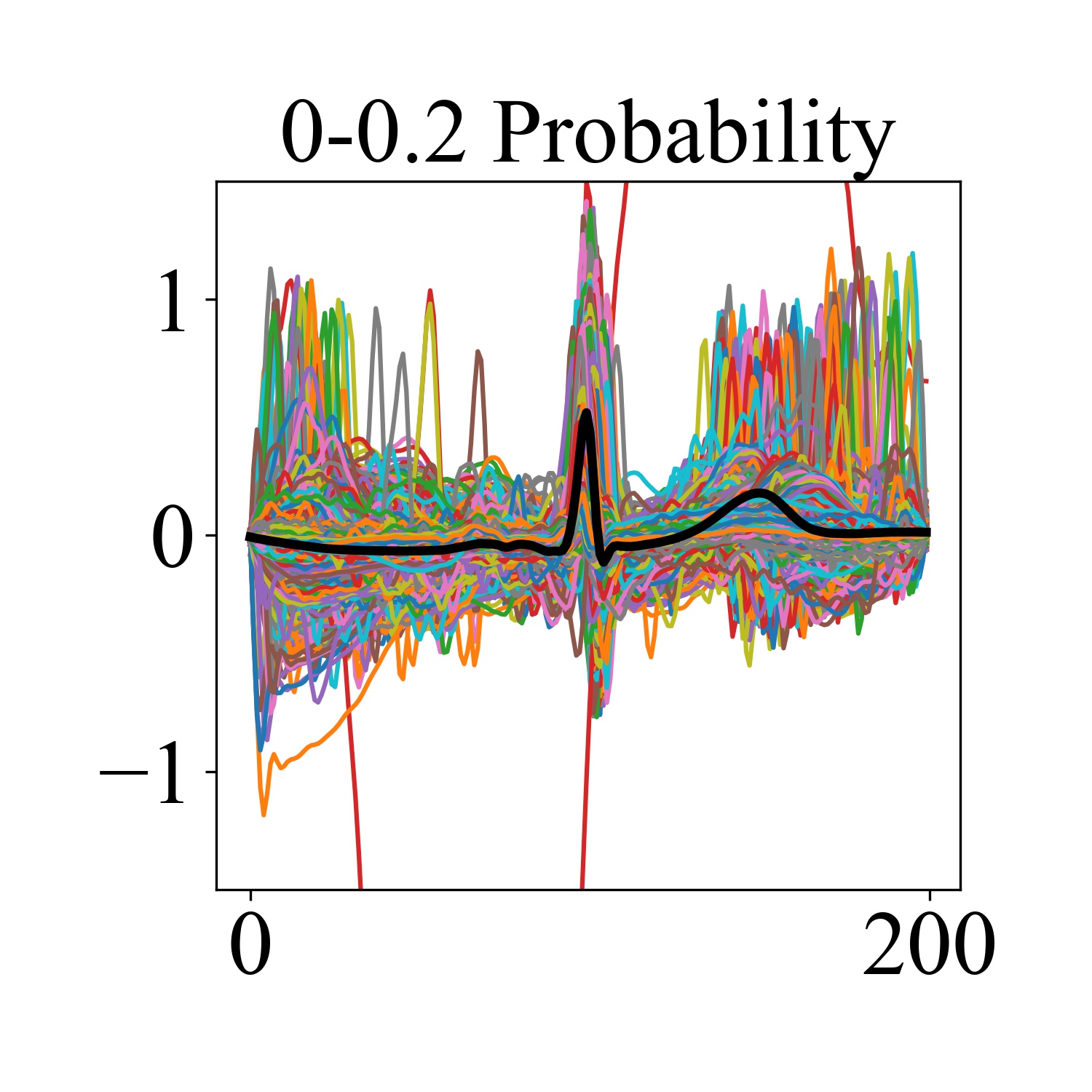}}
\hfill
\subfloat[]{\includegraphics[width=0.2\textwidth]{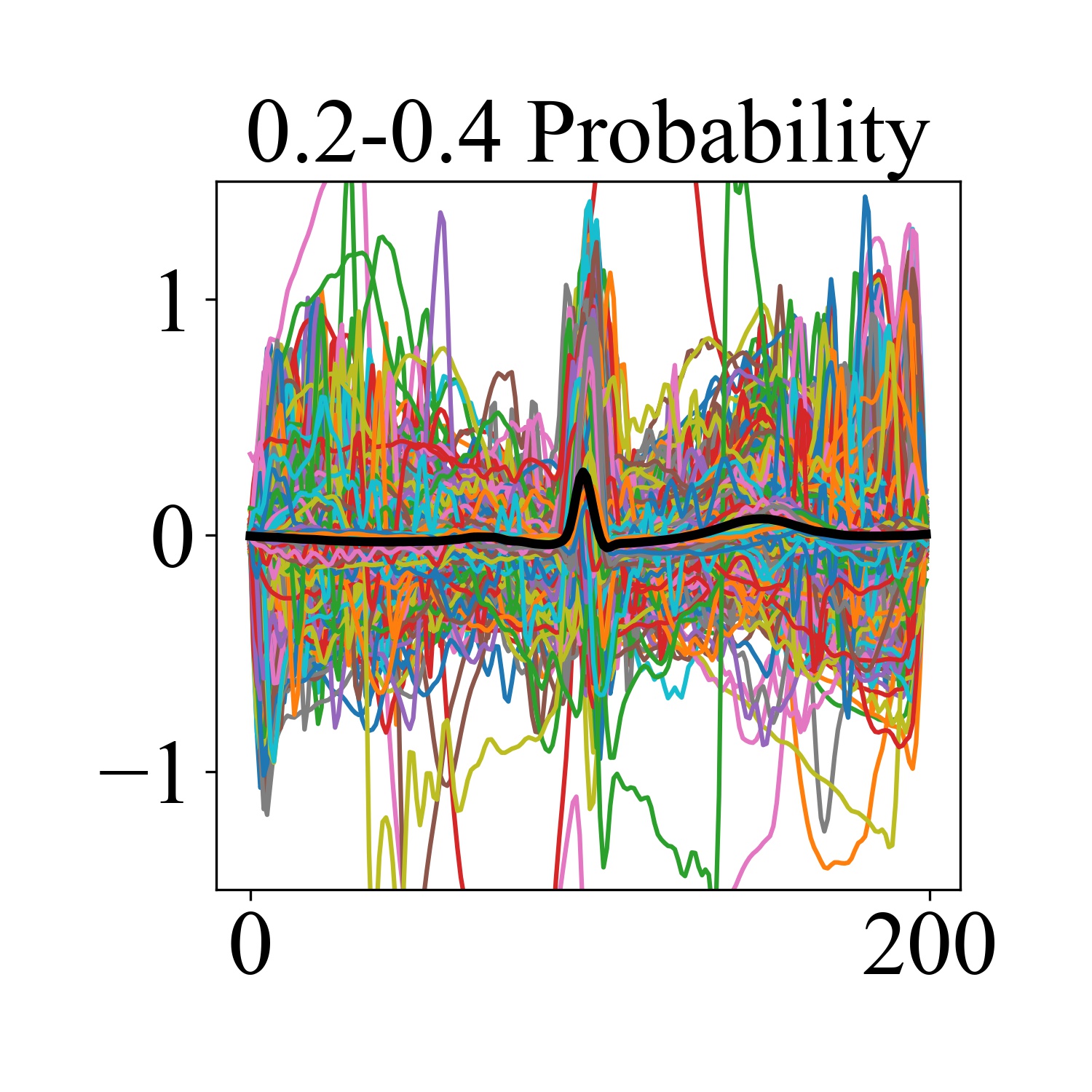}}
\hfill
\subfloat[]{\includegraphics[width=0.2\textwidth]{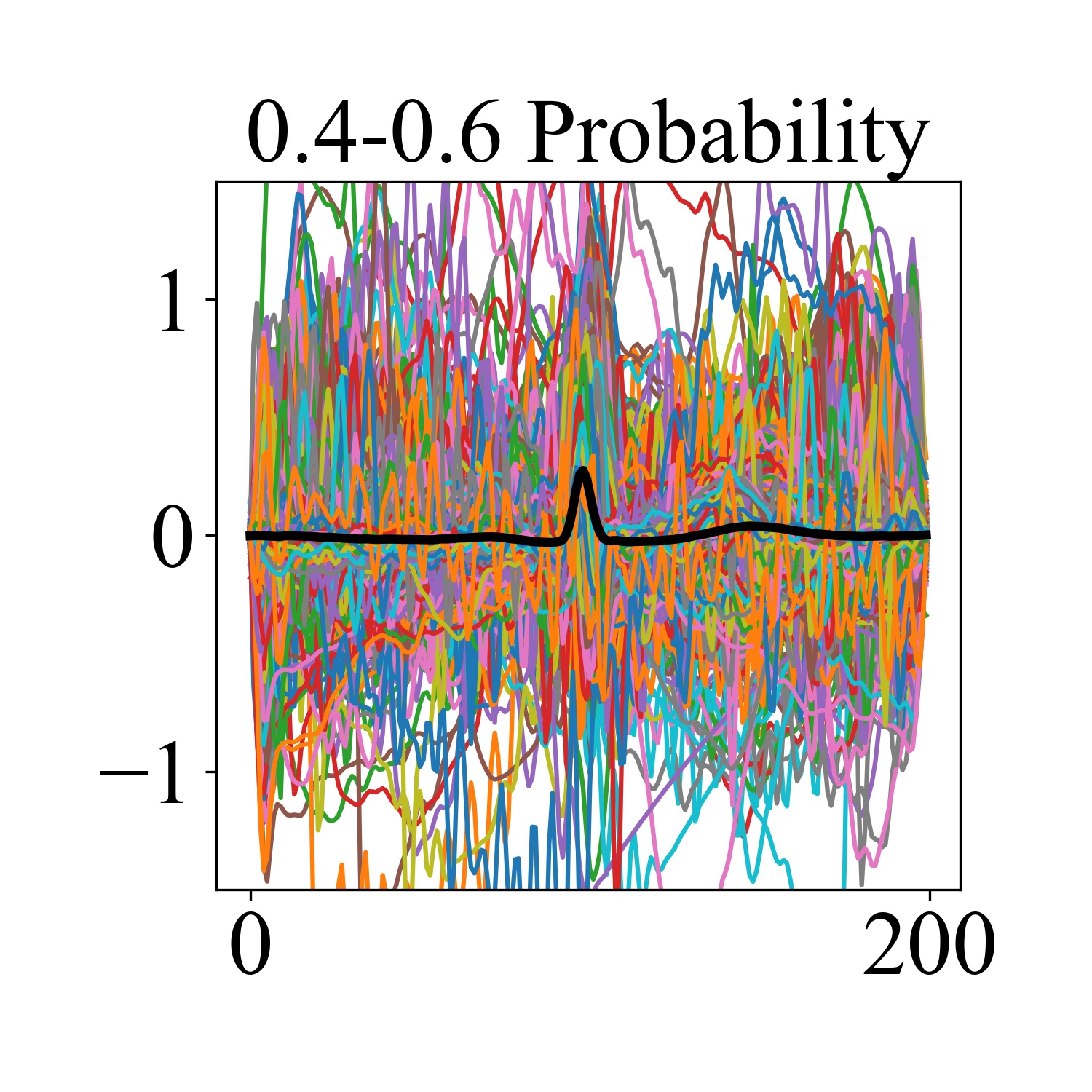}}
\hfill
\subfloat[]{\includegraphics[width=0.2\textwidth]{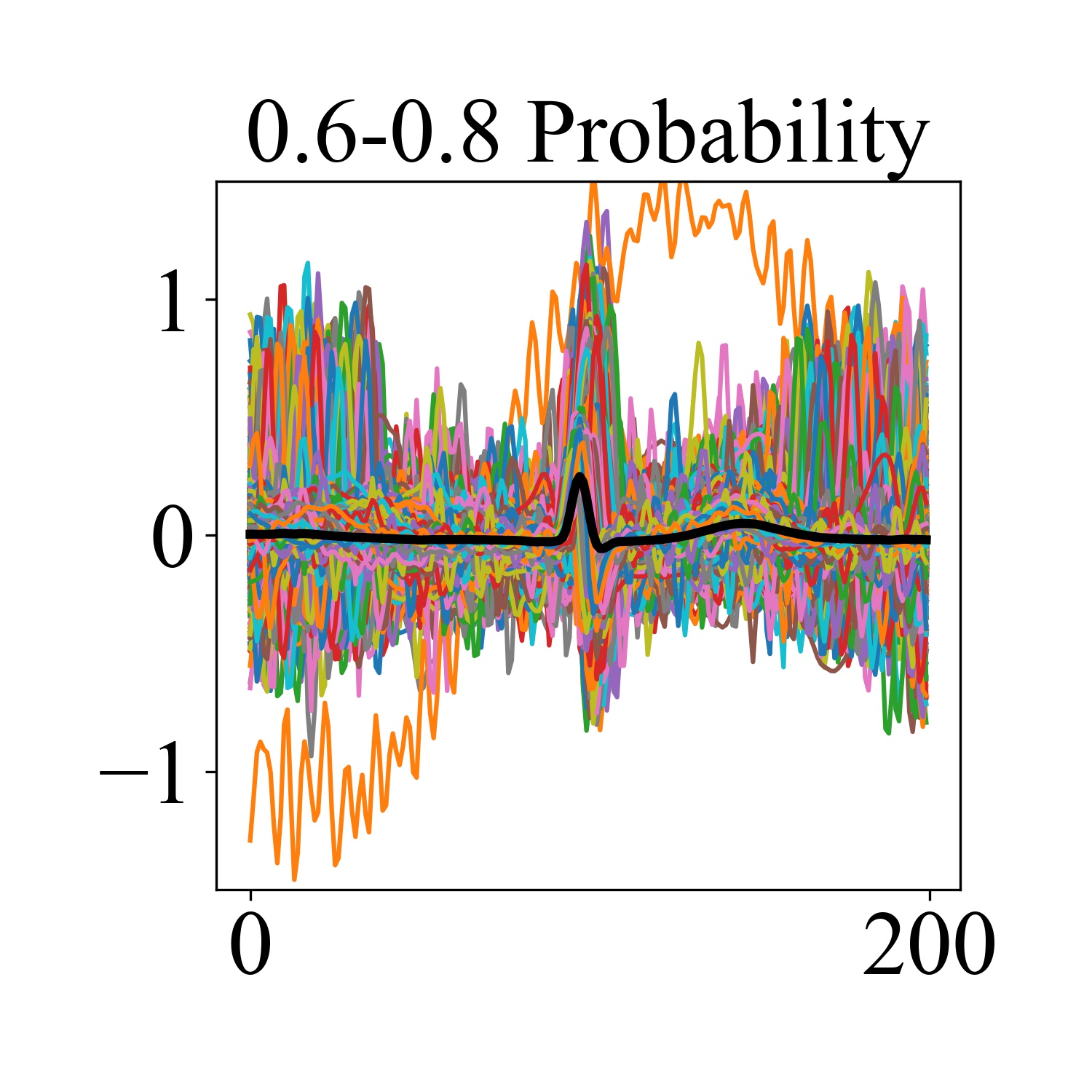}}
\hfill
\subfloat[]{\includegraphics[width=0.2\textwidth]{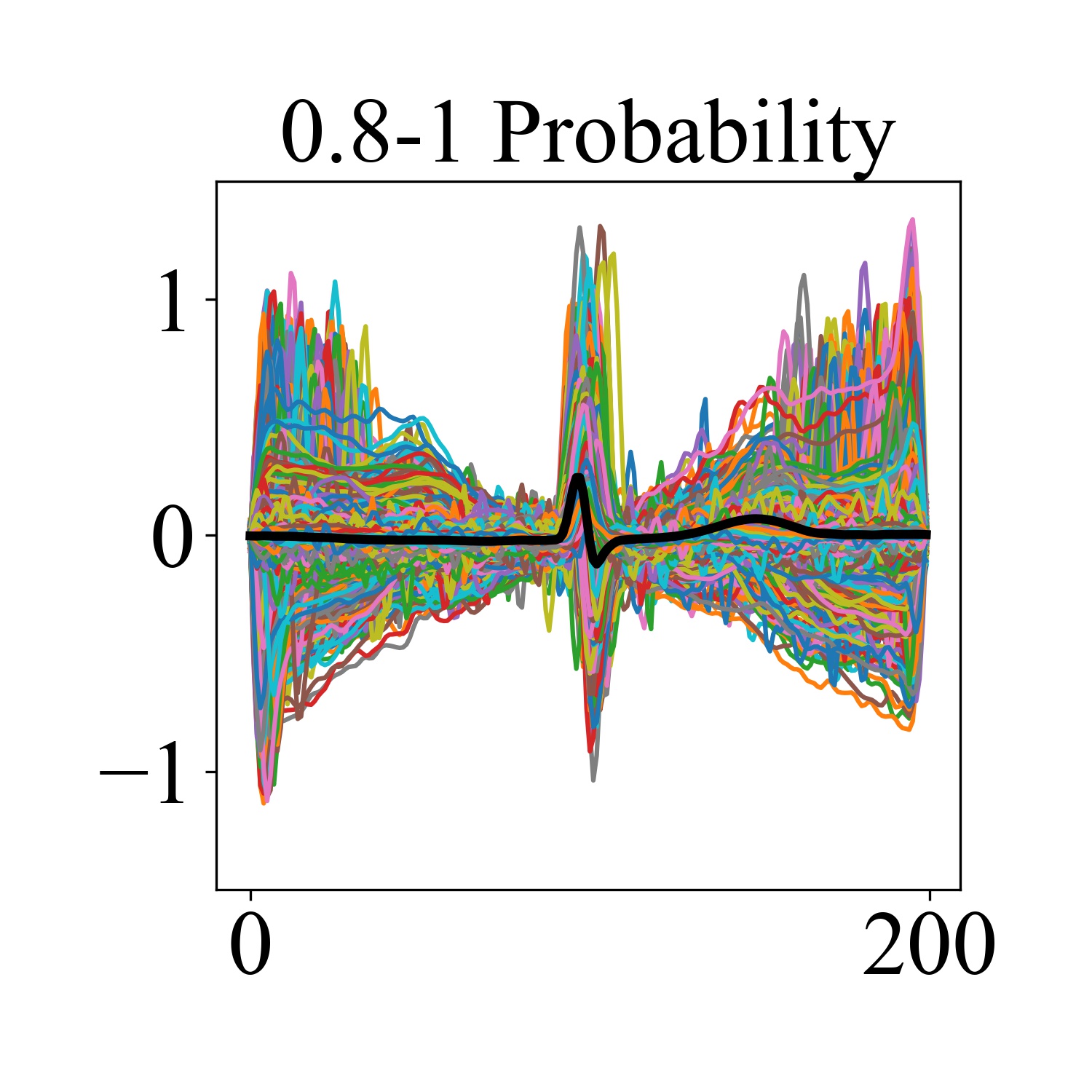}}
\caption{Average sinus beat waveforms for patients at different risk levels. After sorting the probabilities for all patients, they are equally divided into five parts, each representing a risk level. The bold black line represents the average waveform.}
\label{fig:beatShape}
\end{figure*}

\begin{table*}[!ht]
    \centering
    \begin{tabular}{ccccccccc}
    \hline
        \textbf{Class} & \textbf{Index} & \textbf{ACC} & \textbf{REC} & \textbf{PRE} & \textbf{F1} & \textbf{AUC} & \textbf{\# of \miold{Normal} \minew{Sinus}} & \textbf{\# of AF} \\ \hline
        \textbf{} & 1 & 0.5080 & 0.2712 & 0.0791 & 0.1225 & 0.3567 & 285317  (87.33\%) & 41382   (12.67\%) \\ 
        \textbf{} & 2 & 0.4995 & 0.1329 & 0.0470 & 0.0694 & 0.2601 & 216377  (85.96\%) & 35345   (14.04\%) \\ 
        \textbf{Stable} & 3 & 0.7134 & 0.4766 & 0.2605 & 0.3368 & 0.6674 & 307007  (84.73\%) & 55346   (15.27\%) \\ 
        \textbf{} & 4 & 0.7319 & 0.3209 & 0.1425 & 0.1974 & 0.6367 & 156347  (89.73\%) & 17900   (10.27\%) \\ 
        \textbf{} & 5 & 0.6426 & 0.2197 & 0.2243 & 0.2220 & 0.5395 & 115831  (76.80\%) & 34997   (23.20\%) \\ 
        \textbf{} & Avg & 0.6200 & 0.2800 & 0.1500 & 0.1900 & 0.4900 & 216176  (85.39\%) & 36994   (14.61\%) \\ \hline
        \textbf{} & 1 & 0.5423 & 0.3077 & 0.0000 & 0.0001 & 0.3973 & 285317  (100.00\%) & 13      (0.00\%) \\ 
        \textbf{} & 2 & 0.5581 & 0.1652 & 0.0012 & 0.0024 & 0.3078 & 216377  (99.68\%) & 696     (0.32\%) \\ 
        \textbf{Before AF} & 3 & 0.7545 & 0.1104 & 0.0011 & 0.0021 & 0.4415 & 307007  (99.76\%) & 734     (0.24\%) \\ 
        \textbf{} & 4 & 0.7780 & 0.3786 & 0.0042 & 0.0083 & 0.6846 & 156347  (99.76\%) & 383     (0.24\%) \\ 
        \textbf{} & 5 & 0.7651 & 0.2702 & 0.0123 & 0.0235 & 0.5699 & 115831  (98.95\%) & 1225    (1.05\%) \\ 
        \textbf{} & Avg & 0.6800 & 0.2500 & 0.0000 & 0.0100 & 0.4800 & 216176  (99.72\%) & 610     (0.28\%) \\ \hline
        \textbf{} & 1 & 0.5423 & 0.6667 & 0.0001 & 0.0001 & 0.6758 & 285317  (100.00\%) & 12      (0.00\%) \\ 
        \textbf{} & 2 & 0.5578 & 0.1410 & 0.0013 & 0.0025 & 0.2785 & 216377  (99.61\%) & 851     (0.39\%) \\ 
        \textbf{After AF} & 3 & 0.7548 & 0.2014 & 0.0019 & 0.0038 & 0.4933 & 307007  (99.77\%) & 720     (0.23\%) \\ 
        \textbf{} & 4 & 0.7778 & 0.3013 & 0.0033 & 0.0066 & 0.6591 & 156347  (99.75\%) & 385     (0.25\%) \\ 
        \textbf{} & 5 & 0.7648 & 0.2608 & 0.0124 & 0.0236 & 0.5580 & 115831  (98.91\%) & 1277    (1.09\%) \\ 
        \textbf{} & Avg & 0.6800 & 0.3100 & 0.0000 & 0.0100 & 0.5300 & 216176  (99.70\%) & 649     (0.30\%) \\ \hline
        \textbf{} & 1 & 0.5364 & 0.3965 & 0.0350 & 0.0643 & 0.4537 & 285317  (95.98\%) & 11942   (4.02\%) \\ 
        \textbf{} & 2 & 0.5431 & 0.2014 & 0.0214 & 0.0386 & 0.3152 & 216377  (95.44\%) & 10339   (4.56\%) \\ 
        \textbf{BNA} & 3 & 0.7490 & 0.6152 & 0.1182 & 0.1983 & 0.7418 & 307007  (94.95\%) & 16319   (5.05\%) \\ 
        \textbf{} & 4 & 0.7427 & 0.3330 & 0.1176 & 0.1739 & 0.6390 & 156347  (91.87\%) & 13835   (8.13\%) \\ 
        \textbf{} & 5 & 0.7409 & 0.3526 & 0.1044 & 0.1611 & 0.6058 & 115831  (92.94\%) & 8794    (7.06\%) \\ 
        \textbf{} & Avg & 0.6600 & 0.3800 & 0.0800 & 0.1300 & 0.5500 & 216176  (94.64\%) & 12246   (5.36\%) \\ \hline
        \textbf{} & 1 & 0.5460 & 0.7424 & 0.0294 & 0.0566 & 0.6960 & 285317  (98.17\%) & 5327    (1.83\%) \\ 
        \textbf{} & 2 & 0.5647 & 0.7455 & 0.0474 & 0.0892 & 0.7249 & 216377  (97.14\%) & 6365    (2.86\%) \\ 
        \textbf{BNV} & 3 & 0.7593 & 0.9108 & 0.0730 & 0.1352 & 0.9035 & 307007  (97.93\%) & 6478    (2.07\%) \\ 
        \textbf{} & 4 & 0.7796 & 0.7993 & 0.1024 & 0.1815 & 0.8711 & 156347  (96.94\%) & 4932    (3.06\%) \\ 
        \textbf{} & 5 & 0.7612 & 0.6246 & 0.1539 & 0.2469 & 0.7933 & 115831  (93.73\%) & 7744    (6.27\%) \\ 
        \textbf{} & Avg & 0.6800 & 0.7600 & 0.0800 & 0.1400 & 0.8000 & 216176  (97.23\%) & 6169    (2.77\%) \\ \hline
    \end{tabular}
    \caption{The performance of beats within each classification in subgroup analysis. 	extbf{Stable}, representing a given sinus beat without the occurrence of AF in the nearby time period; \textbf{Before AF}, representing a given sinus beat with AF occurring in the preceding time period but not before that; \textbf{After AF}, representing a given sinus beat with AF occurring in the time period before it but not after that; Sinus beat near atrial premature contraction (\textbf{BNA}), representing a sinus beat with the occurrence of atrial premature contraction in the nearby time period; \textbf{BNV}, representing a sinus beat with the occurrence of premature ventricular contraction in the nearby time period. \textbf{\# of \miold{Normal} \minew{Sinus}} represents the number of beats that do not belong to this category, \textbf{\# of AF} represents the number of beats that belong to this category. Time period is 10s.}
    \label{table:subgroup}
\end{table*}

\begin{figure*}[!th]
\centering
\subfloat[]{\includegraphics[width=0.25\textwidth]{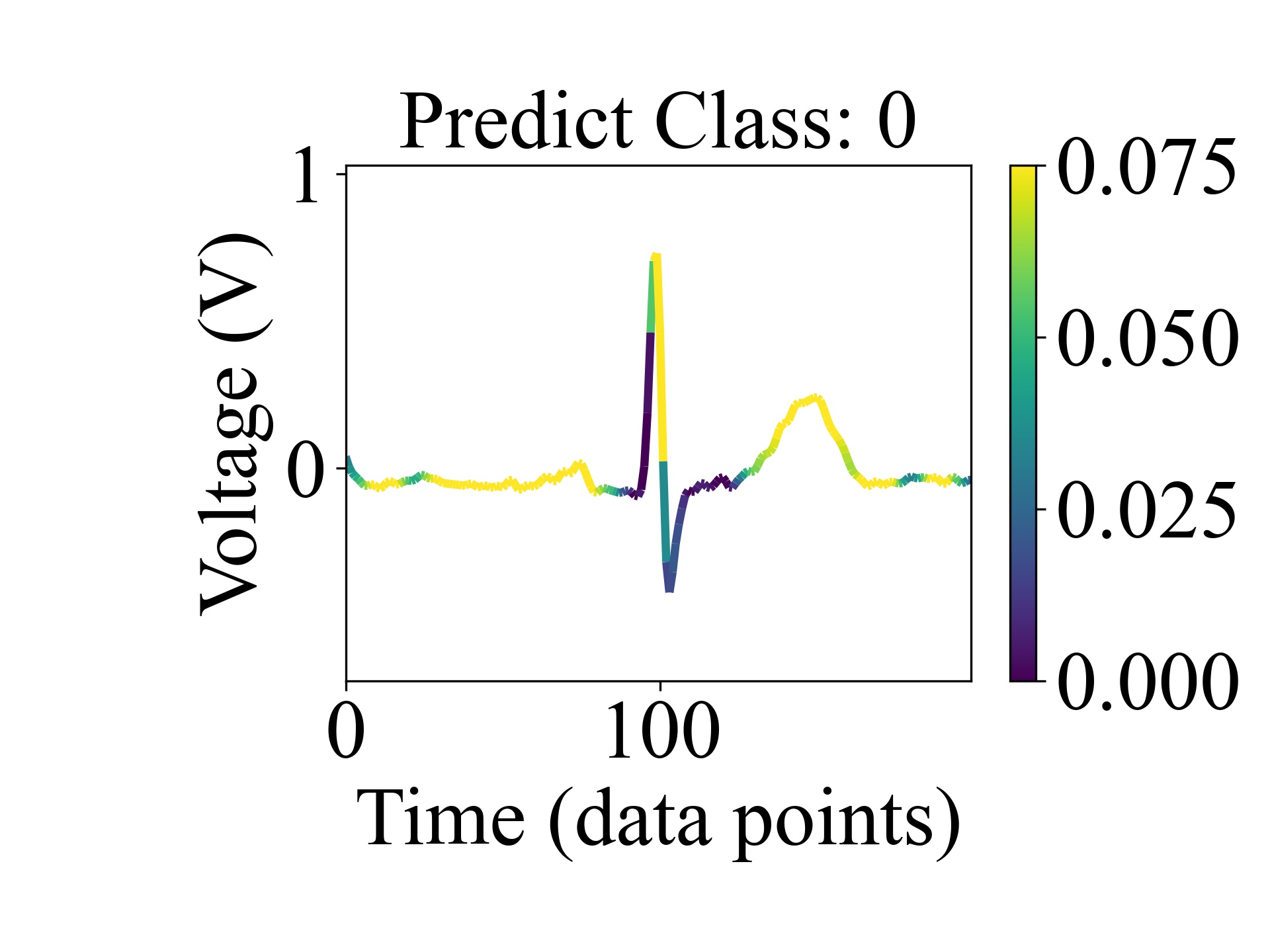}}
\hfill
\subfloat[]{\includegraphics[width=0.25\textwidth]{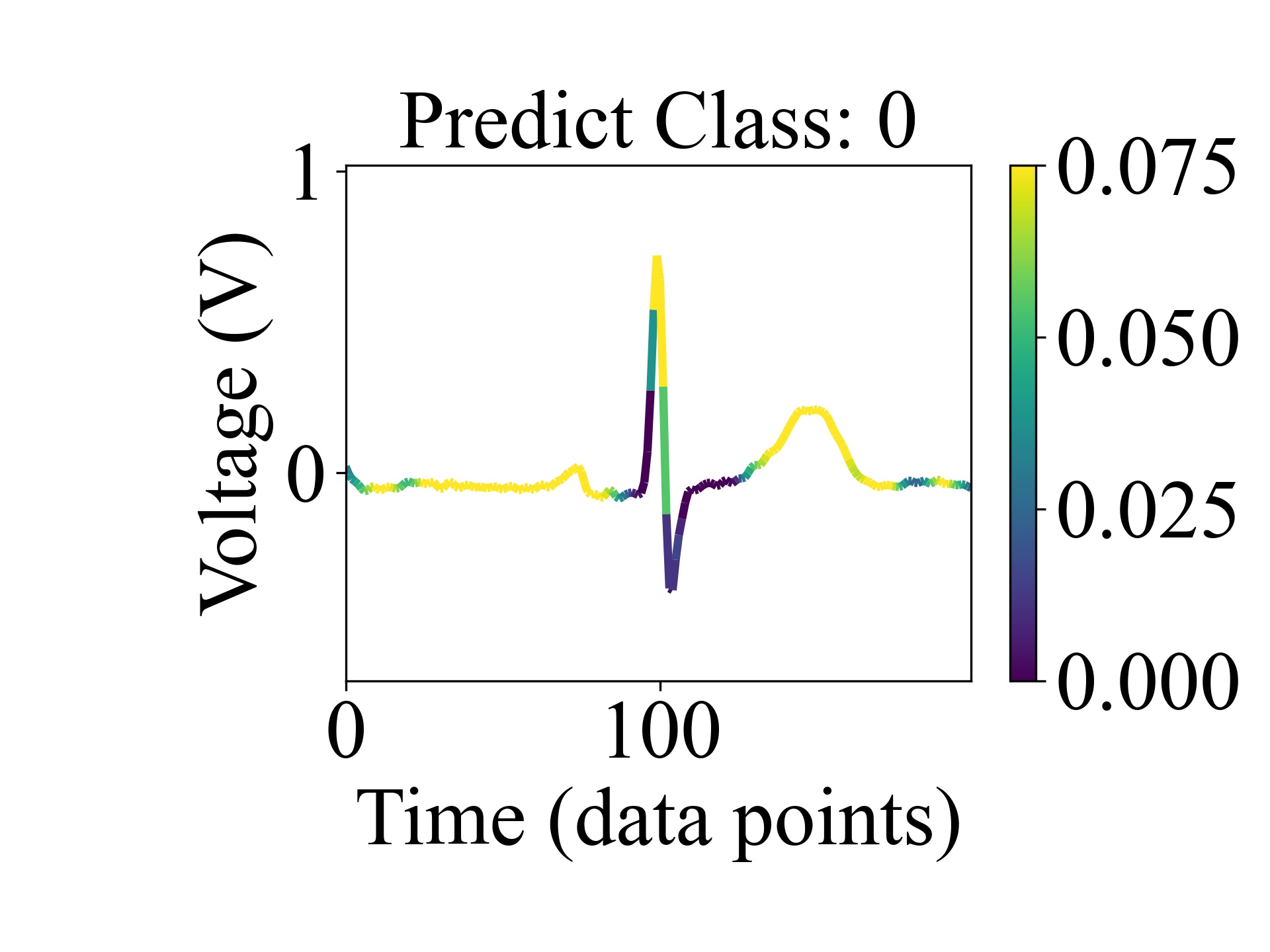}}
\hfill
\subfloat[]{\includegraphics[width=0.25\textwidth]{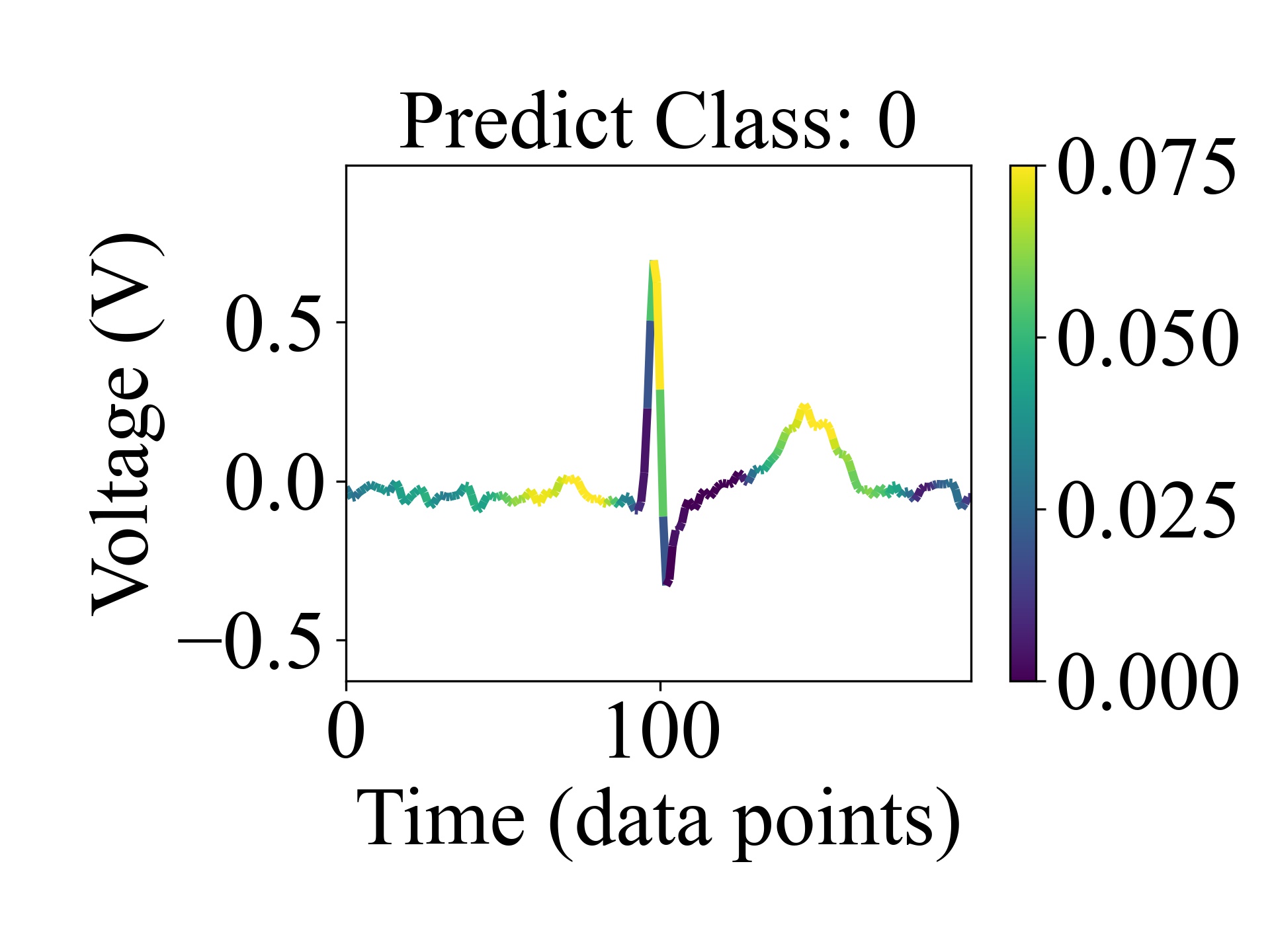}}
\hfill
\subfloat[]{\includegraphics[width=0.25\textwidth]{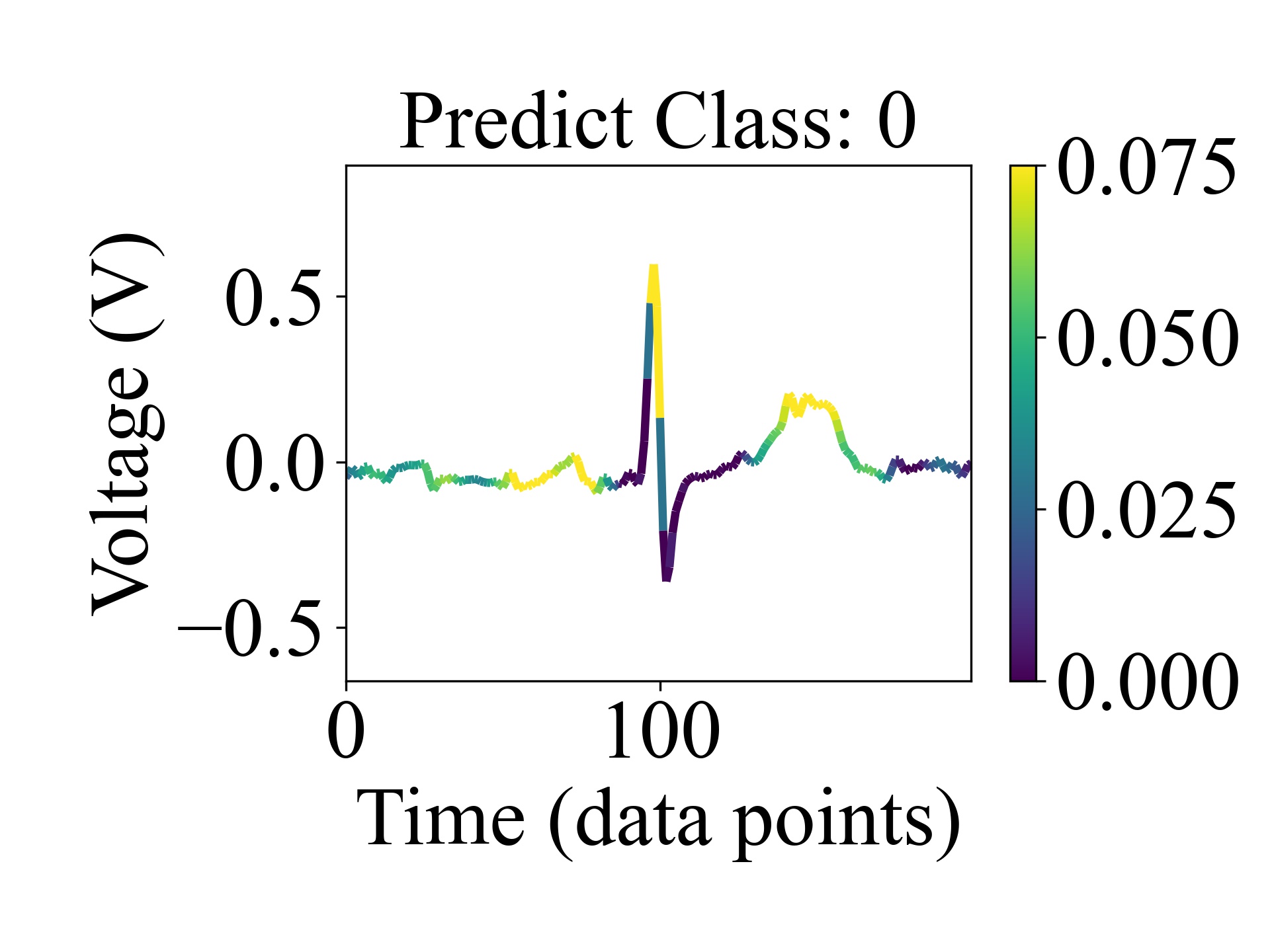}}
\hfill

\subfloat[]{\includegraphics[width=0.25\textwidth]{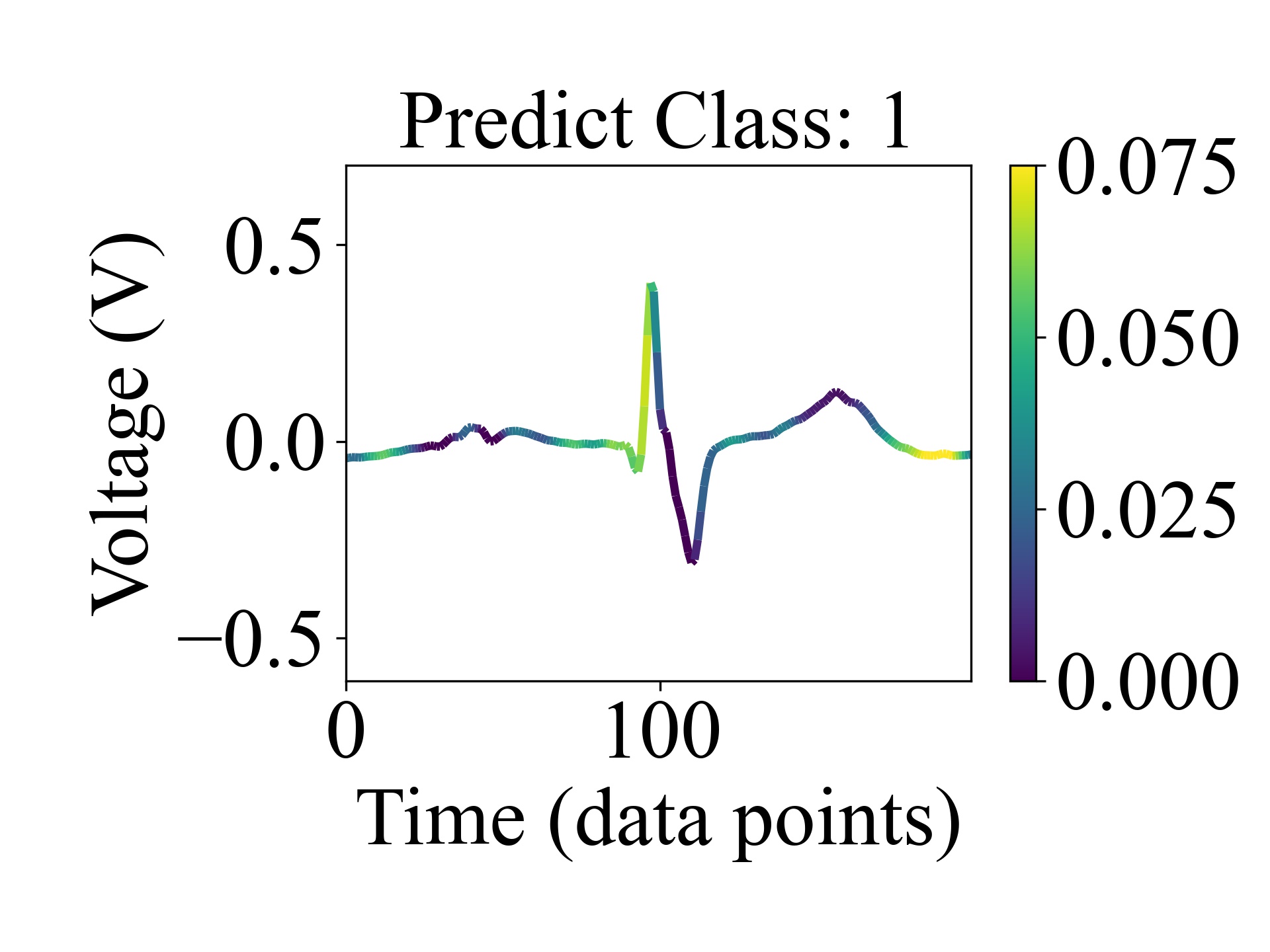}}
\hfill
\subfloat[]{\includegraphics[width=0.25\textwidth]{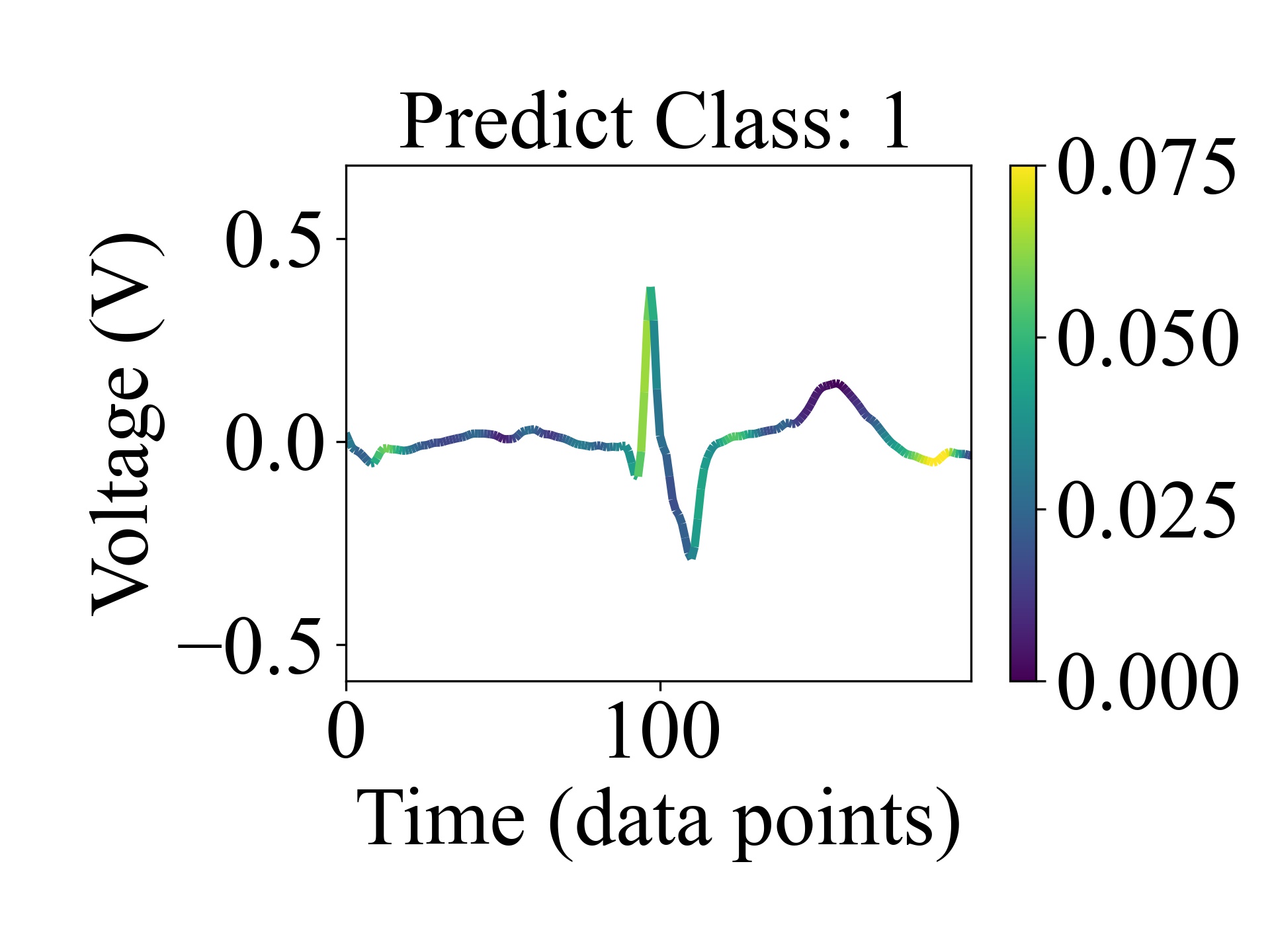}}
\hfill
\subfloat[]{\includegraphics[width=0.25\textwidth]{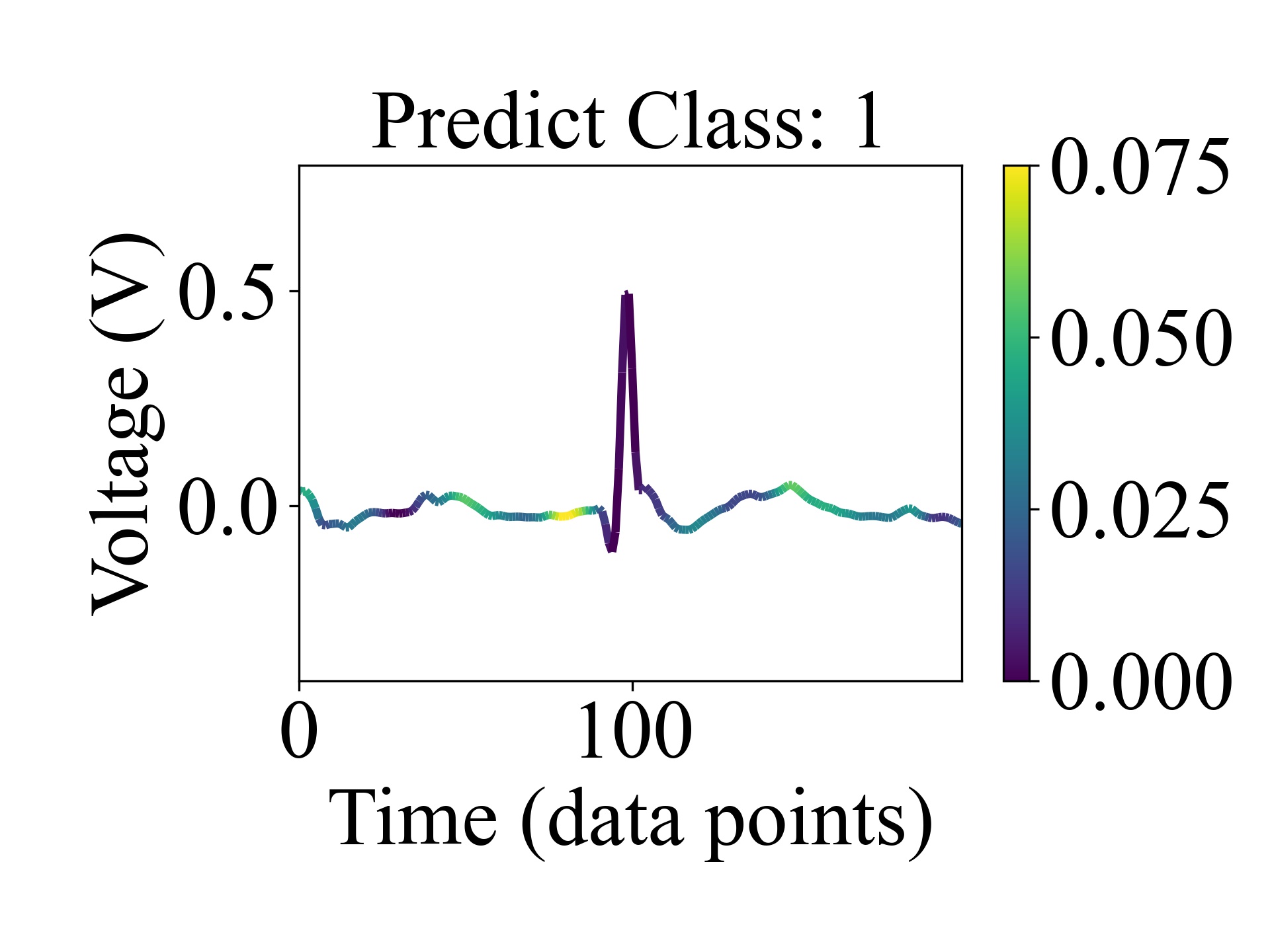}}
\hfill
\subfloat[]{\includegraphics[width=0.25\textwidth]{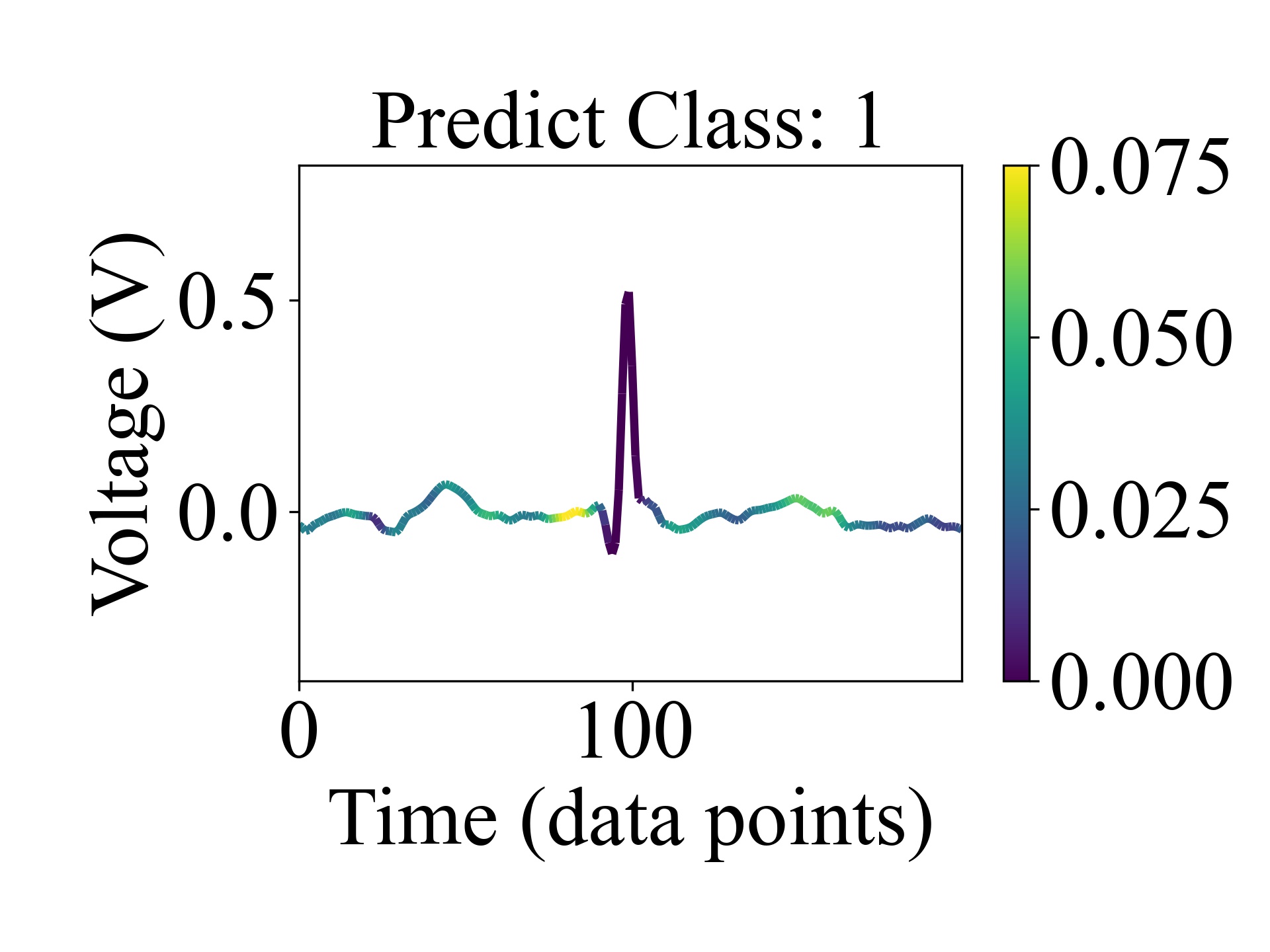}}
\hfill
\caption{The image of CAM. The images (a)-(d) in the upper section represent \miold{normal} \minew{healthy} sinus beats in healthy individuals, while the images (e)-(h) in the lower section represent sinus beats in patients with AF.}
\label{cam}
\end{figure*}

\begin{figure*}[!th]
\centering
\subfloat[]{\includegraphics[width=0.9\textwidth]{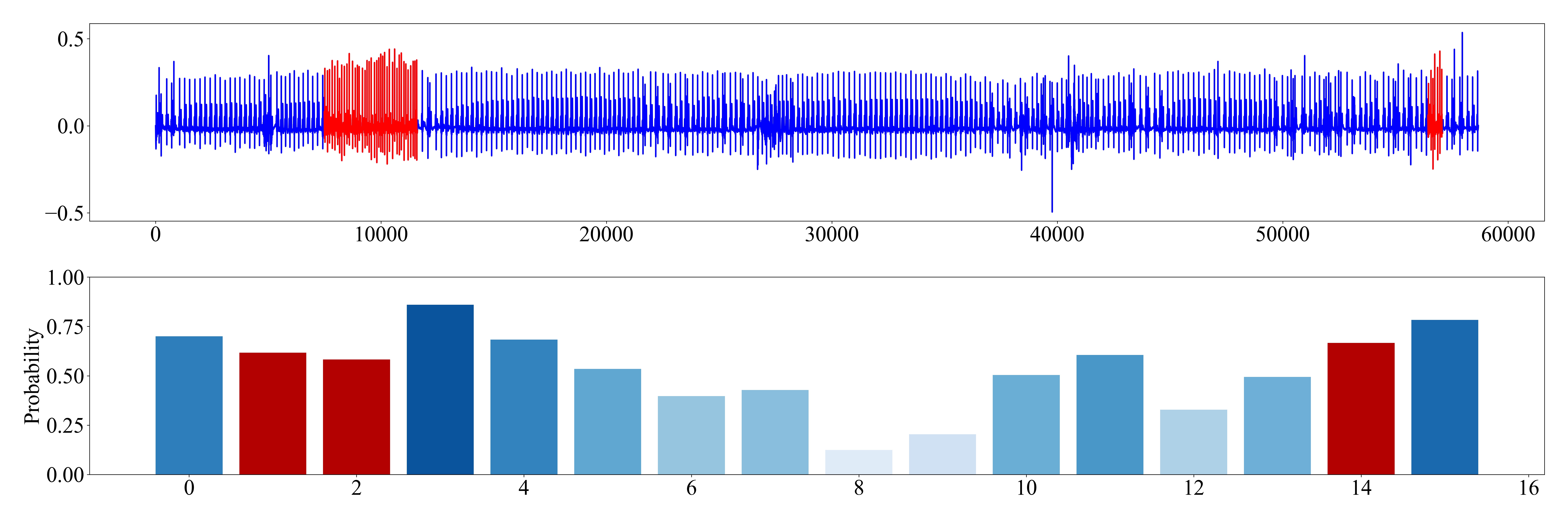}}
\hfill
\centering
\subfloat[]{\includegraphics[width=0.9\textwidth]{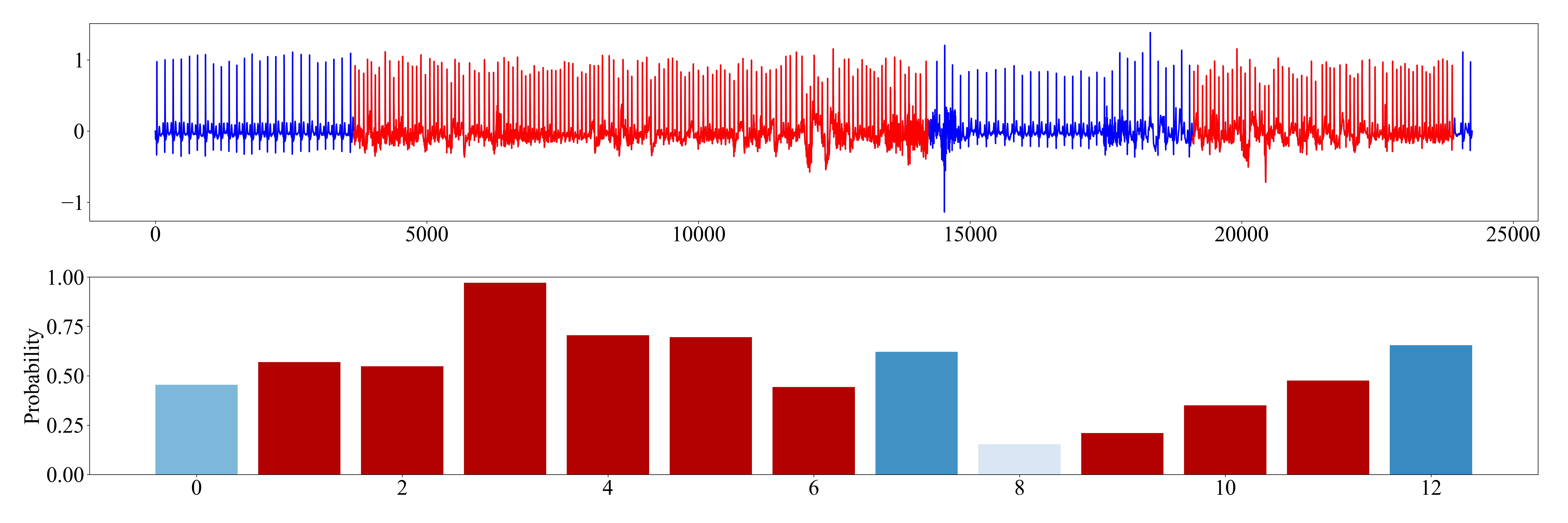}}
\hfill
\caption{Temporal Changes in Average Risk Probability. In Each Subplot, the upper panel depicts an ECG waveform, where the blue segments represent \miold{normal} \minew{sinus} periods, and the red segments indicate AF episodes. The lower panel illustrates the time-averaged risk probability values, with blue denoting \miold{normal} \minew{sinus} segments and red indicating AF segments. Each group consists of 20 beats.}
\label{fig:beatPro}
\end{figure*}

\subsection{Subgroup analysis}

We conducted subgroup analysis on beats in different states. In each model's test results, the data were categorized into AF and non-AF, then the AF patient data were further divided into 5 categories, as detailed in Figure \ref{fig:subgroup}. The non-AF data remained unchanged and were combined with each of the 5 categories separately to observe the final results. The purpose of this combination was to ensure the AUC was displayed correctly.

\minew{Explanation of the 5 different states of Figure \ref{fig:subgroup}:} Stable, representing a given sinus beat without the occurrence of AF in the nearby time period; Before AF, representing a given sinus beat with AF occurring in the preceding time period but not before that; After AF, representing a given sinus beat with AF occurring in the time period before it but not after that; Sinus beat near atrial premature contraction (BNA), representing a sinus beat with the occurrence of atrial premature contraction in the nearby time period; BNV, representing a sinus beat with the occurrence of premature ventricular contraction in the nearby time period. ACC, REC, PRE, F1 and AUC respectively stand for accuracy, recall, precision, and F1 score, area under curve. NoNum represents the number of beats that do not belong to this category, AbNum represents the number of beats that belong to this category, and Ratio represents the ratio of these two quantities.

The experimental results are shown in Table \ref{table:subgroup}. we can observe better model performance in the BNV state, indicating a heightened likelihood of AF occurrence during ventricular arrhythmia. This discovery offers novel diagnostic insights for clinicians. However, the effectiveness of the model decreased in several other scenarios, indicating challenges in detecting AF in other situations.

\begin{table}[!ht]
    \centering
    \begin{tabular}{lll}
    \hline
         & \textbf{Beat} & \textbf{Segment} \\ \hline
        \textbf{Total parameters} & 0.10M & 5.90M \\ 
        \textbf{Training parameters} & 0.10M & 5.89M \\ 
        \textbf{Computational efficiency} & 0.0629s & 0.6784s \\ \hline
    \end{tabular}
    \caption{Comparison of parameter quantity and computational efficiency between beat-level and segment-level models.}
    \label{table:comPara}
\end{table}

\subsection{Model complexity analysis}

In order to illustrate the lightweight nature of the beat-level model, we compare it with the segment-level model used in \cite{attia2019artificial}, both predicting data during a patient's sinus rhythm period. From Table \ref{table:comPara}, we conclude that the beat-level model is more lightweight, making it more suitable for deployment on portable medical devices compared to segment-level models. \minew{Attia et al.'s segment-level model uses 8-lead 10s data, and its model is similar to the Net1D structure used in this paper. Table \ref{table:comPara} shows that the number of model parameters obtained from the 8-lead 10s data is 60 times higher than that of the single-lead 1s model. At the same time, the segment-level model is also about 10 times slower than the beat-level model.}

\subsection{Interpretability and new discoveries}

\paragraph{\textbf{Beat-level Interpretability}}
We used CAM to visualize the interpretability of the model on beat-level data, as shown in Figure \ref{cam}. CAM illustrates the model's attention on beat prediction. Brighter colors indicate higher attention in the corresponding area, while darker colors indicate lower attention. We demonstrated situations where the model predicted high probabilities correctly and compared cases of predicting AF and non-AF. A noticeable observation is that when the data belongs to a patient with AF, the model's correct predictions are based on the presence of a normal P-wave in the beat. In the images predicting 1 (AF), the model did not detect the P-wave, while in the images predicting 0 (non-AF), the model clearly focused on the presence of the P-wave. These findings align with the results of previous studies \cite{giannopoulos2023p, myrovali2023identifying, martinez2013morphological, conte2017usefulness, filos2017beat, tachmatzidis2021beat}, confirming the correctness and accuracy of the model's attention to the data locations.

\paragraph{\textbf{Beat probability trend}}
In the trend risk plot shown in Figure \ref{fig:beatPro}. The upper panel depicts ECG signals, while the lower panel illustrates the average risk probability values for temporal groups. In both cases, blue represents \miold{normal} \minew{sinus} segments, and red indicates AF segments, with each group comprising 20 beats. We observed that beat prediction probabilities increase with the occurrence of AF segments. When sinus beats are around AF segments, the prediction probability is relatively high, indicating that beats around AF segments can be identified by the model. During the \miold{normal} \minew{sinus} stage, where only sinus beats are detected, the prediction probability increases, suggesting that the patient is likely to experience AF in the near future.

\paragraph{\textbf{Average waveforms}}
We present examples of ECG waveforms at different prediction probabilities in Figure \ref{fig:beatShape}. To avoid the influence of the sample size in each interval on the interval waveforms, we sort all beats by prediction probability, and then divide them into five equal parts. These plots provide a visual understanding of the model's predictions. From Figure \ref{fig:beatShape}, we observe that average waveforms can to some extent reflect the health status of the patient's beats. Due to the inherently small size of P-waves, they are not easily observed in the average waveform. However, T-waves, being relatively larger, also gradually disappear with an increase in risk probability. This indicates that when the risk probability is high, the patient's waveform becomes highly unstable, to the extent that both P-waves and T-waves may disappear entirely.

\section{Discussion}

\minew{
In this research, we devised a beat-level algorithm designed to identify AF from ECG signals in sinus rhythm. Our approach involves preprocessing the sinus ECG data to segment it into individual beats, followed by utilizing the Net1D model to learn AF risk from these sinus beats. While our algorithm's AUC performance (0.7591) is lower than the segment-level algorithm AUC reported by Attia et al. (0.87-0.90) \cite{attia2019artificial}, this difference was anticipated. The strength of the beat-level algorithm lies in its detailed analytical capabilities and potential for real-time monitoring, laying the groundwork for future development of more efficient information fusion decision algorithms. In contrast, segment-level algorithms, although performing well on large-scale datasets, may face challenges in terms of real-time applicability and portability in clinical settings.

Regarding the quantity and quality of data, it is essential to highlight the distinction between the single-lead 1-second data utilized in our study and the 8-lead 10-second data employed by Attia et al. \cite{attia2019artificial}. This difference is significant in designing algorithms for portable devices, where lightweight and fast-running algorithms are crucial. Our algorithm demonstrates enhanced adaptability in these aspects, making it particularly suited for extensive screening in resource-limited environments.

The interpretability of the model is crucial for clinical understanding and building trust in AI predictions. Our beat-level interpreter highlights the significant role of the P-wave in AF detection, thereby providing valuable decision support for clinicians. In contrast, the segment-level model proposed by Attia et al. \cite{attia2019artificial} lacks this level of transparency, which may limit clinicians' insight into the model's predictions. Enhancing model interpretability is a key direction for future research.

In the experiment analyzing changes in average risk probabilities, we observed that sinus beats near AF segments exhibit higher risk probabilities. This finding suggests the potential for early warning of AF. Clinicians can prioritize segments with higher risk probabilities during diagnosis and use the BRI for a more detailed analysis.

According to the average waveforms experiment, we found that patients with higher AF risk may have more incomplete heart waveforms. Traditional research focus on the disappearance of P waves, but our research also explores the situation of T waves. The results show that in some ECGs of AF patients, T waves may be lacking or difficult to identify, which may be due to factors such as rapid heart rate and irregular rhythm. The morphology of the T wave can be influenced by a variety of factors, and patients with abnormal T waves have a higher predisposition to AF. AF burden leads to ventricular myocardium remodeling, resulting in T-wave alterations. K{\"o}rtl et al. \cite{kortl2022atrial} observed chromosome-induced cardiomyocytes and found that a higher AF load significantly reduced sarcomere tissue, further supporting our hypothesis. However, we have only observed this phenomenon experimentally, and further research is needed to understand the underlying physiological characteristics.

Finally, we acknowledge the limitations of using only the CPSC2021 dataset, including the simplification of label assignment and the potential for overclassification, as well as the constrained number of patients. In this study, we employed a simplified binary label assignment method, categorizing patients based solely on the presence or absence of AF episodes. This approach may lead to overclassification, especially when distinguishing between transient or infrequent AF episodes. Since the majority of patients exhibit a predominantly sinus rhythm in their ECG signals, this simplified label assignment might result in misclassifications, particularly in cases where AF episodes are not prominently manifested. These limitations could impact the model's learning capacity and influence experimental outcomes. To address these challenges, we plan to expand the dataset in future research to include a more diverse range of patient populations and refine the dataset partitioning strategy.
}

\section{Conclusion and Future Work}
In conclusion, we have established a beat-level algorithm for identifying the risk of AF in distinguishing ``sinus rhythm in patients with AF'' and ``sinus rhythm in \miold{normal} \minew{healthy} individuals''. \minew{Specifically, we extract multiple 1 second beat-level data from sinus ECG and input them into the Net1d model to obtain the risk probability for each beat.} \minew{The construction of Net1d is straightforward and flexible, allowing each module to independently adjust its hyperparameters, thus enabling scalable expansion of both the model's depth and width.} We proposed a BRI, BID and TRI, along with several findings, showcasing meaningful clinical value. Providing timely AF risk reports to patients, enhancing collaboration between physicians and AI through the interpretability of model results, facilitates the prompt identification of AF, allowing for early intervention and treatment of this condition.

In the future, we aim to improve the model's generalization by incorporating more high-quality datasets. Additionally, we plan to explore more rational labeling strategies based on the number or duration of AF episodes to enhance the model's performance.

\section*{Acknowledgments}

The authors gratefully acknowledge the financial supports by the National Natural Science Foundation of China \minew{(under Grant 62202332, Grant 62102008 and Grant 62176183) and the University Student Innovation and Entrepreneurship Training Program (under Grant 202410060104)}.




 \bibliographystyle{elsarticle-num} 
 \bibliography{elsarticle-template-num}








\end{document}